# Biofilm self-patterning: mechanical forces drive a reorientation cascade

Japinder Nijjer[1], Changhao Li[2], Qiuting Zhang[1], Haoran Lu[1], Sulin Zhang[2,3*], Jing Yan[1,4]*

[1]Department of Molecular, Cellular and Developmental Biology, Yale University, New Haven, CT, USA.

[2]Department of Engineering Science and Mechanics, Pennsylvania State University, University Park, PA, USA.

[3]Department of Biomedical Engineering, Pennsylvania State University, University Park, PA, USA.

[4]Quantitative Biology Institute, Yale University, New Haven, CT, USA.

**Abstract:**

In growing active matter systems, a large collection of engineered or living autonomous units metabolize free energy and create order at different length scales as they proliferate and migrate collectively. One such example is bacterial biofilms, which are surface-attached aggregates of bacterial cells embedded in an extracellular matrix. However, how bacterial growth coordinates with cell-surface interactions to create distinctive, long-range order in biofilms remains elusive. Here we report a collective cell reorientation cascade in growing *Vibrio cholerae* biofilms, leading to a differentially ordered, spatiotemporally coupled core-rim structure reminiscent of a blooming aster. Cell verticalization in the core generates differential growth that drives radial alignment of the cells in the rim, while the radially aligned rim in turn generates compressive stresses that expand the verticalized core. Such self-patterning disappears in adhesion-less mutants but can be restored through opto-manipulation of growth. Agent-based simulations and two-phase active nematic modeling reveal the strong interdependence of the driving forces for the differential ordering. Our findings provide insight into the collective cell patterning in bacterial communities and engineering of phenotypes and functions of living active matter.

**Main Text:**

The spatiotemporal patterning of cells is a fundamental morphogenetic process that has profound effects on the phenotypes and functions of multicellular organisms[1–3]. In the prokaryotic domain, bacteria are often observed to form organized multicellular communities surrounded by extracellular matrices[4,5], known as biofilms[6,7], which are detrimental due to persistent infections, clogging of flows, and surface fouling, but can be beneficial in the context of wastewater treatment[8] and microbial fuel cells[9]. During development, biofilms exhibit macroscopic morphological features ranging from wrinkles, blisters, to folds[10–12]. At the cellular scale, recent progress in single-cell imaging has revealed the reproducible three-dimensional architecture and developmental dynamics of biofilms[13–16]. However, how the cellular ordering emerges from individual bacterium trajectories remains poorly understood. In particular, it remains unclear how cell proliferation is coordinated with intercellular interactions in a growing biofilm to elicit robust self-patterning against bacteria's inherent tendency to grow in an unstructured manner[17–19]. An understanding of how individual cell growth links to collective patterning as a result of self-generated forces can provide insights into the developmental program of biofilms[6], their physical properties[20], and the engineering of living and nonliving active-matter analogs[21,22].

To bridge the gap between interactions at the cellular scale and patterns at the community scale, here we combine single-cell imaging and agent-based simulations to reveal the underlying mechanism for self-patterning in biofilm formed by *Vibrio cholerae*, the causal agent of the pandemic cholera. We observe that biofilm-dwelling bacteria self-organize into an aster pattern, which emerges from a robust reorientation cascade, involving cell verticalization in the core and radial alignment in the growing rim. We reveal that the verticalized core generates directional flow that drives radial alignment of the cells in the periphery, while the radially aligned rim generates compressive stresses that expand the verticalized core, leading to a robust, inter-dependent differential orientational ordering. Based on these findings, we derive a two-phase active nematic model for biofilm self-patterning, which is potentially generalizable to other developmental systems with growth-induced flows[23,24]. Our findings suggest that the self-generated cellular force landscape, rather than chemical signaling or morphogen gradients as often seen in eukaryotic cells[25], controls pattern formation in biofilms.

## ***V. cholerae*** **biofilms self-organize into aster patterns**

We imaged the growth of *V. cholerae* biofilms confined between glass and an agarose gel at single-cell resolution (Fig. 1a). We used a constitutive biofilm producer locked in a high c-di-GMP state[26] and focused on the biophysical aspects of self-organization. To simplify our studies, we focused on a mutant missing the cell-to-cell adhesion protein RbmA – this strain is denoted as WT* – although our analysis is equally applicable to strains with cell-to-cell adhesion (Extended Data Fig. 1). Using confocal microscopy, the 3D architecture of the biofilms was captured over time from single founder cells to mature biofilms consisting of thousands of cells (Fig. 1b; Supplementary Video 1). An adaptive thresholding algorithm was used to segment individual cells in the 3D biofilm (Extended Data Fig. 2; Supplementary Information Section 1) from which the location and direction of each rod-shaped bacterium were identified (Fig. 1c-f). Strikingly, cells in the basal layer of WT* biofilms *reproducibly* self-organized into an aster pattern, consisting of a core with tilted or "verticalized" cells and an outward splaying rim with radially aligned cells (Fig. 1d; Extended Data Fig. 1).

We quantified the degree of cell ordering in the basal layer using a radial order parameter[27], $S = 2\langle(\hat{\boldsymbol{n}}_\parallel \cdot \hat{\boldsymbol{r}})^2\rangle - 1$, where $\hat{\boldsymbol{n}}_\parallel$ is the projection of the cell direction on the basal plane and $\hat{\boldsymbol{r}}$ is the unit vector along the radial direction (Fig. 1g). $S$ equals 1 for cells that are aligned in an aster, $-1$ for cells that are aligned in a vortex, and 0 for cells that are randomly oriented. We found that cells in WT* biofilms exhibited a reproducible tendency to align radially ($S = 0.54 \pm 0.07$) in the rim. Since previous work has shown that cell-to-surface adhesion controls overall biofilm morphology[12,14,28], we hypothesized that cell-to-surface adhesion mediates the dynamic core-rim patterning of the biofilm. To test this hypothesis, we deleted the genes encoding cell-to-surface adhesion proteins Bap1 and RbmC[29–31] ($\Delta BC$) and found that the radial order was destroyed in the resulting biofilms and cells assumed random orientations in the basal plane with $S = 0.11 \pm 0.11$ (Fig. 1e, g). Concomitant with the disorder was the absence of a verticalized core; most cells in the basal layer were parallel to the substrate. We further confirmed the important role of cell-to-surface adhesion by titrating *rbmC* expression: increasing cell-to-surface adhesion enhanced the self-patterning, resulting in more verticalized cells and stronger radial alignment (Fig. 1h; Extended Data Fig. 3). Furthermore, removing the extracellular matrix by deleting the key *Vibrio* polysaccharide ($\Delta vpsL$)[32] resulted in locally aligned microdomains of horizontal cells without long-range order ($S = 0.02 \pm 0.08$; Fig 1f, g), in line with previous studies on growing 2D bacterial colonies[18,19]. These observations suggest that exopolysaccharide production controls a *local* order-to-disorder transition, whereas cell-to-surface adhesion controls a *global* order-to-disorder transition.

To determine the driving forces behind the observed orientational ordering, we extended a previous agent-based model,[33] taking into account cell-to-cell and cell-to-surface interactions (Supplementary Information Section 2). Our agent-based modeling reproduced the observed aster pattern formation in adherent cells, but not in nonadherent cells, in agreement with experiments (Extended Data Fig. 4; Supplementary Video 2). As the agent-based model only incorporates mechanical interactions, without any biochemical signals, our results suggest that the emergent patterns originate primarily from the mechanical interplay between the cells and between cells and the substrate.

## Surface adhesion drives ordering through differential growth

In molecular liquid crystals, a lower temperature favors order due to the entropic driving force. For out-of-equilibrium systems, such as growing biofilms, the driving force for ordering is more complex. We hypothesized that radial organization arises from the mechanical coupling between cells through their self-generated flow field[34], inspired by the alignment of rod-shaped objects under fluid shear[35]. Note that biofilm-dwelling cells are nonmotile; flow in this context is generated through cell growth and cell-cell interactions. To test our hypothesis, we tracked cell orientations and trajectories *simultaneously* during biofilm development by using strains expressing a single intracellular punctum (Fig. 2a-e, Extended Data Fig. 5; Supplementary Video 3)[16]. As WT* biofilms grew, cells towards the center tilted away from the substrate, developing a core of verticalized cells that expanded over time (Fig. 2c). The resulting growth-induced flow field had a zero-velocity core (Fig. 2a, d), corresponding to the verticalized cells that project their offspring into the third dimension (Fig. 1d). In contrast, in the nonadherent mutant, the velocity field simply scaled linearly with the radial position. From the measured velocity field, we extracted the *apparent* in-plane proliferation rate $g$ (Fig. 2b, Fig. 2d inset). We found that $g$ was uniform in the nonadherent biofilm: all cells in the basal layer were predominantly parallel to the substrate and therefore contributed to the basal layer expansion. In contrast, in the WT* biofilm, a growth void ($g \approx 0$) emerged in the center, with nearly uniform growth in the outer growing rim. Concomitant with the initiation of differential growth, cells aligned in an aster pattern, marked by a growing $S(r)$ with a rising peak near the edge of the verticalized core (Fig. 2e).

## A reorientation cascade governs biofilm self-patterning

We hypothesize that a mechanical synergy between cell verticalization, growth-induced flow, and aster pattern formation propels a reorientation cascade for biofilm self-patterning. On one hand, cell-to-surface adhesion coupled with growth-induced mechanical stresses leads to *stably* anchored, verticalized cells in the biofilm center, which results in differentially oriented proliferation. One the other, differential proliferation drives cellular flows that radially align the cells in the rim, which in turn facilitates cell verticalization and core expansion. Below, we analyze the dynamic interplay of these two reorientation processes.

**Step 1:** To illustrate the formation and stabilization of the verticalized core, we consider a reduced problem consisting of a spherocylindrical cell that is parallel and adhered to a substrate and squeezed by two neighbors (Supplementary Information Section 3). The resulting energy landscape displays two distinct mechanical instabilities (Fig. 2f). The first instability corresponds to the verticalization event reported earlier[33,36–38]. Briefly, cells in a growing population mechanically push each other, generating pressure. This pressure accumulates and eventually exceeds a threshold, causing cells to rotate away from the substrate (verticalize). The second instability corresponds to the "pinch-off" of these verticalized cells. In this case, neighboring cells generate forces in the out-of-plane direction, causing ejection of the verticalized cells from the substrate. For WT* cells, our analysis shows that pinching a vertical cell off the surface is energetically much more costly than verticalizing a horizontal cell. Therefore, pinch-off is kinetically hindered and verticalized cells can *stably* inhabit the basal layer. The smaller the cell-to-surface adhesion, the smaller the energy difference between the two instabilities (Extended Data Fig. 6) and therefore, the less stable the verticalized cells. The energy difference vanishes in nonadherent cells, resulting in spontaneous ejection of mutant cells upon verticalization. This explains the absence of verticalized cells in the mutant biofilms and bacterial colonies (Fig. 1e, f). In the WT* biofilms, verticalization preferentially occurs near the center where pressure is relatively high, leading to an expanding verticalized core[33]. Since rod-shaped cells grow and divide along their long axes, this spatial segregation of cell orientation leads to spatially patterned differential growth.

**Step 2:** Next, we employ active nematic theory[34,39,40] to elucidate how differential growth can induce radial alignment. Defining the nematic order parameter $\boldsymbol{Q} = 2\langle \hat{\boldsymbol{n}}_{\parallel} \otimes \hat{\boldsymbol{n}}_{\parallel} - \boldsymbol{I}/2 \rangle$ as the head-tail symmetric tensor of cell orientation, mesoscopically averaged over a small region, its evolution in a surrounding flow $\boldsymbol{u}$ is given by[41]

$$(\partial_t + \boldsymbol{u} \cdot \nabla)\boldsymbol{Q} - \Gamma \boldsymbol{H} = \lambda \boldsymbol{E} + \boldsymbol{\omega} \cdot \boldsymbol{Q} - \boldsymbol{Q} \cdot \boldsymbol{\omega}, \tag{1}$$

where the right-hand side quantifies the driving force for the rod-shaped particles to rotate within a velocity gradient field. Here $\boldsymbol{E} = \frac{1}{2}[\nabla \boldsymbol{u} + \nabla \boldsymbol{u}^T - (\nabla \cdot \boldsymbol{u})\boldsymbol{I}]$ is the traceless strain-rate tensor, $\boldsymbol{\omega} = \frac{1}{2}(\nabla \boldsymbol{u} - \nabla \boldsymbol{u}^T)$ is the vorticity tensor, and $\lambda$ is the flow-alignment parameter. For rod-shaped objects $\lambda > 0$, corresponding to a tendency for the rods to align with flow streamlines[17]. Finally, the nematic alignment term $\Gamma \boldsymbol{H}$ relaxes $\boldsymbol{Q}$ towards a bulk state with minimal angular variation, however, its contribution in biofilms is expected to be negligible since cells are buffered from each other by soft exopolysaccharides (Supplementary Information Section 4). Assuming axisymmetry, the evolution of the cell orientation field is given by[19,34]

$$\partial_t \Theta + u_r \partial_r \Theta = -f(r,t)\sin(2\Theta), \tag{2}$$

where $\Theta$ is the angle between the local orientation field and the radial direction, $f = (\lambda r/4q)\partial_r(u_r/r)$ quantifies the aligning torque due to gradients in the flow field, and $q$ quantifies the degree of local ordering (Supplementary Information Section 4). From $\partial_t \Theta \sim -f \sin(2\Theta)$, we find that a nonzero $f$ causes cells to rotate, and the direction of rotation is critically dependent on the sign of $f$.

Unlike passive liquid crystals, biofilm-dwelling cells generate their own velocity field through growth. Assuming uniform density, mass conservation requires $\nabla \cdot \boldsymbol{u} = g(r)$. In nonadherent mutant biofilms and bacterial colonies, growth is exclusively in-plane with a uniform growth rate $\gamma$, resulting in a linear velocity field, $u_r = \gamma r/2$, and thus a vanishing driving force for cell alignment ($f = 0$). Under this condition, cells are simply advected outwards without any tendency to align, leading to a disordered pattern. In contrast, in WT* biofilms, verticalization stabilizes an expanding in-plane growth void, $r_0(t)$. This corresponds to a differential growth rate $g(r)$: 0 for $r \leq r_0$ and $\gamma$ for $r > r_0$. The resulting velocity field is $\gamma(r - r_0^2/r)/2$ for $r > r_0$, leading to a *strictly positive* driving force for radial alignment, $f = \frac{\lambda \gamma r_0^2}{4qr^2} > 0$, in the outer growing rim. In this case, $\Theta$ dynamically approaches 0, characteristic of an aster (Extended Data Fig. 7). In fact, long-range order can be induced whenever a 2D growing bacterial colony deviates from an isotropically expanding pattern, for instance when confined in a rectangular geometry[42,43] or during inward growth[34]. This model thus reveals that differential growth, established by a verticalized core ($r_0 \neq 0$), generates the driving force for radial alignment in a growing biofilm. This driving force vanishes in the absence of a core ($r_0 = 0$), leading to a disordered phenotype.

**Imposing a growth void reproduces radial ordering**

A key prediction of the active nematic theory is that a growth void is *sufficient* to induce radial organization. To test this prediction, we patterned a growth void into an otherwise disordered biofilm. Specifically, we started with a nonadherent biofilm already grown for 17 hours and used a 405 nm laser to selectively kill the cells in the center. The vestiges of the dead cells sustained a growth void (Extended Data Fig. 8), mimicking the verticalized core in the WT* biofilm. Consistent with our model prediction, the proliferating cells aligned radially over time in biofilms with a growth void, whereas biofilms without a growth void remained disordered (Fig. 3a-c). Conversely, our theory predicts that *excess* growth at the biofilm center should lead to $f < 0$ and therefore to vortex formation (Supplementary Information Section 4). Indeed, in another set of experiments, we observed that the nonadherent cells aligned circumferentially when excess growth was introduced at the center (Extended Data Fig. 9). We also quantitatively tested the validity of the model by prescribing a growth void with a fixed size $r_0$ in a set of simplified 2D agent-based models (Extended Data Fig. 7; Supplementary Video 4). We found that the instantaneous angular velocity of individual cells scaled linearly with $\sin(2\Theta)/r^2$ and increasing $r_0$ led to a quadratic increase in the angular velocity, all in agreement with the theory (Fig. 3d). Note that in both simulations and experiments, the radial order quickly saturated in the patterned biofilm with a fixed $r_0$, since the aligning force decays with $1/r^2$ as cells are advected outward. Thus, a growing $r_0(t)$ is necessary to reinforce radial alignment during biofilm expansion. This is indeed the case in WT* biofilms: growth of the outer rim accumulates pressure to generate more verticalized cells and expand the verticalized core, which in turn continuously drives alignment in the outer horizontal cells. To interrogate the mechanical interplay between these reorientation processes, we next develop a minimal physical model coupling verticalization of individual cells to the long-range radial ordering.

**Two-phase model of cell organization**

We decompose the biofilm into populations of two phases with vertical and horizontal cells and take the phase fractions to be $\rho$ and $1-\rho$, respectively. The growth kinetics of the phases are governed by

$$\begin{aligned} \partial_t \rho + \nabla \cdot (\boldsymbol{u}\rho) &= C(p)(1-\rho), \\ \partial_t (1-\rho) + \nabla \cdot \big(\boldsymbol{u}(1-\rho)\big) &= \gamma(1-\rho) - C(p)(1-\rho). \end{aligned} \tag{3a,b}$$

Here we assume that the horizontal-to-vertical conversion is driven by the local pressure $p$, where $C(p)$ is the conversion rate. We further assume that pressure arises from friction with the substrate $\nabla p = \eta \boldsymbol{u}$, where $\eta$ is the friction coefficient, and that only the horizontal cells generate growth in the basal layer, $\nabla \cdot \boldsymbol{u} = \gamma(1-\rho(\boldsymbol{r}))$. Combined with Eq. (2), these equations generate a complete continuum description of the dynamics of cell growth and reorientation in biofilms (Supplementary Information Section 4). Numerical solutions of the model quantitatively reproduce the cascade of self-organization events (Fig. 4a-d), showing the intimate spatiotemporal coupling between cell verticalization and radial alignment.

Many salient features of the experimental results are recapitulated by the model: for example, $S(r)$ reaches a maximum near the verticalized core where the driving force is the strongest. Interestingly, the model reveals a frozen core where cells cease to reorganize (compare Fig. 2e and Fig. 4d): as the in-plane velocity goes to zero, the driving force to rotate also vanishes – cells in the core are thus locked as a "fossil record" that memorizes the mechanical history they have experienced. Importantly, the model yields robust results: regardless of the initial conditions and choice of parameters (Extended Data Fig. 10), a WT* biofilm always patterns itself following the sequence shown in Fig. 4e. Our two-phase active nematic model thus elucidates the reproducible mechanical blueprint that guides biofilm development.

## Discussion

To conclude, our results reveal a mechanically driven self-patterning mechanism in bacterial biofilms in which cells synergistically order into an aster pattern. Specifically, we showed that surface adhesion leads to stable cell verticalization, which in turn directs radial cell alignment during surface expansion. Evidently, this inter-dependent differential ordering involves biofilm-wide, bidirectional mechanical signal generation and transmission, in contrast to the biochemical signaling widely observed in other living organisms. In *On Growth and Form*[44], D'Arcy Thompson wrote: "… growth [is] so complex a phenomena…rates vary, proportions change, and the whole configuration alters accordingly." Although over a century old, this statement still rings true today. Our two-phase active nematic model provides a mathematical formalism for this statement in the context of bacterial biofilms.

Spontaneous flow generation is a common phenomenon in various developmental systems, including zebrafish embryonic development[24], ventral furrow formation in *Drosophila*[23], etc. While flow causes bulk morphological changes in these systems, in biofilms it acts to transmit mechanical forces and drive long-range organization. It is intriguing to contemplate whether the synchronous mechanical coupling between differentially grown cells and the resulting pattern could be generalized to other organisms with anisotropic growth of polarized cells. In a broader context, cell polarity and organization critically underlie collective cell function and normal development, as exemplified by topological defects that mediate 2D-to-3D transitions in motile bacterial colonies[45] and cell death and extrusion in epithelial layers[46]. Our findings hence shed light on the biomechanical control of cell organization through the spatiotemporal patterning of growth and pave the way to controlling cell organization by encoding synthetic biological circuits or optogenetic manipulation[47].

## Methods

Bacterial strains and cell culture

Strains used in this study were derivatives of the *V. cholerae* strain C6706 containing a missense mutation in the *vpvC* gene (*vpvC*$^{W240R}$), which resulted in constitutive biofilm production through the upregulation of c-di-GMP (rugose/Rg strain). For the majority of the results presented in this work, we used a strain in which the gene encoding the cell-to-cell adhesion protein RbmA was deleted to minimize the effects of intercellular adhesion; however, we found that our analysis equally applied to the rugose strain (Extended Data Fig. 1). We primarily worked with two other mutants: 1) Δ*BC* which included additional deletions of *bap1* and *rbmC* genes, and 2) Δ*vpsL* in which a key exopolysaccharide biogenesis gene was deleted in the rugose background (RgΔ*vpsL*). In the absence of *bap1* and *rbmC*, the Δ*BC* mutant cells were unable to adhere to the substrate (referred to as the nonadherent mutant throughout the text). In the absence of *vpsL*, the cells did not properly synthesize exopolysaccharides and consequently, all accessary matrix proteins, which bind to the exopolysaccharide, did not function properly[14]. For velocity field measurements, we used strains containing the μNS protein from the avian reovirus fused to an mNeonGreen fluorescent protein, which were shown to self-assemble into a single intracellular punctum[16,48]. All strains used in the study were also modified to constitutively produce either mNeonGreen or mScarlet-I fluorescent proteins. Mutations were genetically engineered using either the pKAS32 exchange vector[49] or the MuGENT method[50]. For a full list of strains used, see Table S1. Biofilm growth experiments were performed using M9 minimal media supplemented with 0.5% glucose (w/w), 2 mM $MgSO_4$, 100 μM $CaCl_2$, and the relevant antibiotics as required (henceforth referred to as M9 media).

Experiments began by first growing *V. cholerae* cells in liquid LB overnight under shaken conditions at 37°C. The overnight culture was back-diluted 30× in M9 media and grown under shaken conditions at 30°C for 2-2.5 hours until it reached an $OD_{600}$ value of 0.1-0.2. The regrown culture was subsequently diluted to an $OD_{600}$ of 0.001 and a 1 µL droplet of the diluted culture was deposited in the center of a glass-bottomed well in a 96-well plate (MatTek). Concurrently, agarose was dissolved in M9 media at a concentration of 1.5-2% (w/V) by microwaving until boiling and then placed in a 50°C water bath to cool without gelation. After cooling, 200 nm far-red fluorescent particles (Invitrogen F8807) were mixed into the molten agarose at a concentration of 1% (V/V) to aid in image registration. Next, 20 µL of the molten agarose was added on top of the droplet of culture and left to cool quickly at room temperature, to gel, and to trap the bacterial cells at the gel-glass interface. Subsequently, 100 µL of M9 media was added on top of the agarose gel, serving as a nutrient reservoir for the growing biofilms. The biofilms were then grown at 30°C and imaged at designated times.

Image acquisition

Images were acquired using a confocal spinning disk unit (Yokogawa CSU-W1), mounted on a Nikon Eclipse Ti2 microscope body, and captured by a Photometrics Prime BSI CMOS camera. A 100× silicone oil immersion objective (N.A. = 1.35) along with 488 nm, 561 nm and 640 nm lasers were used for imaging. This combination of hardware resulted in an *x*-*y* pixel size of 65 nm and a *z*-step of 130 nm was used. For end-point imaging, biofilms were imaged after 12-24 hours of growth and only the 488 nm channel, corresponding to the mNeonGreen expressing cells, was imaged. For time-lapse imaging, samples were incubated on the microscope stage in a Tokai Hit stage top incubator while the Nikon perfect focus system was used to maintain focus. Images were captured every 30 minutes, and in addition to the 488 nm channel, the 640 nm channel was used to image the fluorescent nanoparticles.

For velocity measurements, cells constitutively expressing mScarlet-I and mNeonGreen-labelled puncta were imaged using a slightly modified procedure. The 488 nm channel, corresponding to the puncta, was imaged every 2-10 minutes while the 561 nm channel, corresponding to the cells, was imaged every 1-2 hours. This procedure allowed us to image the relatively bright puncta with low laser intensity and therefore minimal photobleaching and phototoxicity, as high temporal resolution is required to accurately track puncta motion. To further reduce photobleaching and phototoxicity, we used a $z$-step of 390 nm when imaging the puncta. When imaging the cells, a $z$-step of 130 nm was used in the mScarlet-I channel to sufficiently resolve the 3D position and orientation of the cells. We also restricted our attention to the basal flow field and therefore only imaged the bottom 3 μm of each biofilm. All images shown are raw images rendered by Nikon Elements software unless indicated otherwise.

Overview of image analysis

Raw images were first deconvolved using Huygens software (SVI) using a measured point spread function. The deconvolved three-dimensional confocal images were then binarized, layer by layer, with a locally adaptive Otsu method. To accurately segment individual bacterium in the densely packed biofilm, we developed an adaptive thresholding algorithm. For more details see Supplementary Information Section 1. Once segmented, we extracted the cell positions by finding the center of mass of each object, and the cell orientations by performing a principal component analysis. The positions and directions of each cell were converted from cartesian $(x, y, z, \hat{n}_x, \hat{n}_y, \hat{n}_z)$ to cylindrical polar $(r, \psi, z, \hat{n}_r, \hat{n}_\psi, \hat{n}_z)$ coordinates where the origin was found by taking the center of mass of all of the segmented cells in the $(x, y)$ plane. We define the out-of-plane component of the direction vector as $n_\perp = \hat{\boldsymbol{n}} \cdot \hat{\boldsymbol{z}}$ and the in-plane component as $\boldsymbol{n}_\parallel = \hat{\boldsymbol{n}} -$ $(\hat{\boldsymbol{n}} \cdot \hat{\boldsymbol{z}})\hat{\boldsymbol{z}}$, which we normalize as $\hat{\boldsymbol{n}}_{\parallel} = \boldsymbol{n}_{\parallel}/|\boldsymbol{n}_{\parallel}|$. Reconstructed biofilm images were rendered using Paraview.

Measurement of the growth-induced velocity field

To measure the growth-induced velocity field we used particle tracking velocimetry on the puncta trajectories. The deconvolved puncta images were first registered using Matlab built-in functions. Puncta were then detected by first identifying local intensity maxima in the 3D images, and sub-pixel positional information was found by fitting a parabola to the pixel intensity around the maxima. This procedure was repeated for all frames yielding puncta locations over time which were then connected from frame to frame using a standard particle-tracking algorithm[51]. The radial velocity $u_r$ was calculated by fitting a straight line through the time vs. radial displacement data over a time interval of 1 hr.

Opto-manipulation of cell growth

Previous work has shown the bactericidal effects of high energy near-UV light[52]; therefore, we used spatially patterned 405 nm light to kill a subset of cells within a biofilm. Specifically, an Opti-Microscan XY galvo-scanning stimulation device with a 405 nm laser was used to selectively illuminate and kill cells within a cylindrical region at the center of the biofilm. We verified cell killing by staining the sample with propidium iodide (Extended Data Fig. 8). The same procedure used to measure the growth-induced velocity field (see above) was applied to the irradiated biofilm and the control to measure cell orientation and trajectory dynamics simultaneously.

3D agent-based simulations

Building on the agent-based simulations developed by Beroz *et al.*[33] and others[37,53,54], we modelled cells as spherocylinders with a cylinder of length $L(t)$ and two hemispherical caps of radius $R$. The growth of each cell was assumed to be unidirectional and exponential, where the growth rate $\gamma$ was normally distributed with a mean of $\gamma_0$ and a standard deviation of $0.2\gamma_0$. Here noise was added to account for the inherent stochasticity in cell growth and division. Each cell elongated exponentially until its length reached $L_{max} = 2L_0 + 2R$, at which point it was replaced by two daughter cells with the length $L_0$. The doubling time can be calculated to be $t_{\text{double}} = \frac{1}{\gamma}\log\left(\frac{10R+6L_0}{4R+3L_0}\right)$. The cell-to-cell and cell-to-substrate contact mechanics were described by linear elastic Hertzian contact mechanics[55], with a single contact stiffness $E_0$; note that $E_0$ corresponds to the modulus of the soft exopolysaccharide in the matrix (~ $10^2$ Pa) rather than the cell itself, which is much stiffer (~ $10^5$ Pa). Correspondingly, the $R$ value we used (0.8 μm) is larger than the physical size of a cell (~ 0.4 μm). The parameter values we used were calibrated by rheological measurement and microscopy analysis, and have been shown to successfully capture the dynamics of biofilm-dwelling cells in prior work[33]. The cell-to-substrate adhesion energy was assumed to be linear with the contact area, with adhesion energy density $\Gamma_0$. We incorporated two viscous forces to represent the motion of biofilm-dwelling cells at low Reynold's number: 1) a bulk viscous drag for all degrees of freedom, and 2) a much larger in-plane surface drag for cells near the substrate, representing the resistance to sliding when a cell is adhered to the substrate via the surface adhesion proteins RbmC/Bap1. The two damping forces also ensured that the cell dynamics were always in the overdamped regime.

We treated the confining hydrogel as a homogenous, isotropic, and linear elastic material using a coarse-grained approach. The geometry of the coarse-grained gel particles was assumed to be spherical with a radius $R_{\text{gel}}$. The interaction between particles was modeled using a harmonic pairwise potential and a three-body potential related to bond angles. The contact repulsions between the gel particles and the cells as well as between the gel particles and the substrate were described using linear elastic Hertzian contact mechanics. We treated the adhesion between the gel and the substrate using a generalized JKR contact model[56] and we also included a small viscous damping force to ensure the dynamics remained overdamped. The hydrogel was initialized by annealing the system to achieve an amorphous configuration.

Simulations were initialized with a single cell lying parallel to the substrate and surrounded by gel particles. Initially, a small hemispherical space surrounding the cell was vacated to avoid overlap between the cell and the hydrogel particles. We fixed a small number of hydrogel particles near the boundaries to provide anchoring for the elastic deformation of the hydrogel; however, the boundaries were kept sufficiently far away from the biofilm to minimize any boundary effects. We applied Verlet integration and Richardson integration to numerically integrate the equations of motion for the translational and rotational degrees of freedom, respectively. We implemented the model based on the framework of LAMMPS[57], utilizing its built-in parallel computing capability. For a more detailed description on the ABS, see Supplementary Information Section 2.

Quasi-2D agent-based simulations

To further verify the alignment dynamics of the continuum model quantitatively (Eq. 2, Main Text), we developed a set of quasi-2D simulations to mimic the laser irradiation experiments. To simplify the system, the translational and rotational degrees of freedom related to the vertical direction were ignored, while all other parameters were kept the same as the 3D simulations. In each simulation, the bacteria first proliferate normally for 12 hrs, at which point the growth rate of the cells within a radius $r_0$ from the center of the biofilm was set to 0, mimicking the zone of dead cells caused by laser irradiation (Extended Data Fig. 7). In agreement with experiments, the simulated biofilm was initially randomly oriented ($S \approx 0$); however, cells tended toward an aster pattern and $S$ increased over time when the growth void was introduced. The predicted rate at which the cells were driven towards this pattern, in the Lagrangian frame of reference of the cells, is $D_t\Theta = -\frac{\lambda\gamma r_0^2}{4qr^2}\sin(2\Theta)$. We tested this relationship in the simulation data by comparing the instantaneous angular velocity $D_t\Theta$ and $\frac{1}{r^2}\sin(2\Theta)$ (Fig. 3d). Note that we nondimensionalized the $x$-axis by the final colony radius 25 μm. We varied the radius of the growth void $r_0$ and repeated the same procedure and for each simulation run, we plotted the slope of the line of best fit versus $r_0^2$ (Fig. 3d inset).

Data and materials availability

Matlab codes for single-cell segmentation are available online at Github: https://github.com/Haoran-Lu/Segmentation_3D-processing/releases/tag/v1.0. Other data are available upon request.

**Acknowledgments:** We thank Drs. A. Mashruwala and Y. Xu for their help in the initial experiments. We thank Drs S. Mao, T. Cohen, and J.-S. Tai for helpful discussions and B. Reed and M. Zhao for help with developing the ABSs.

**Author contributions:** J.N. and J.Y. designed and performed the experiments. J.N., Q.Z., H.L., and J.Y. analyzed data. C.L. and S.Z. developed the agent-based simulations. J.N. developed the continuum theory. J.N., C.L., S.Z., and J.Y. wrote the paper.

**Competing interests:** The authors declare that they have no competing interests.

**Supplementary Materials:** Supplementary information is available for this paper.

Correspondence and requests for materials should be addressed to either jing.yan@yale.edu or suz10@psu.edu

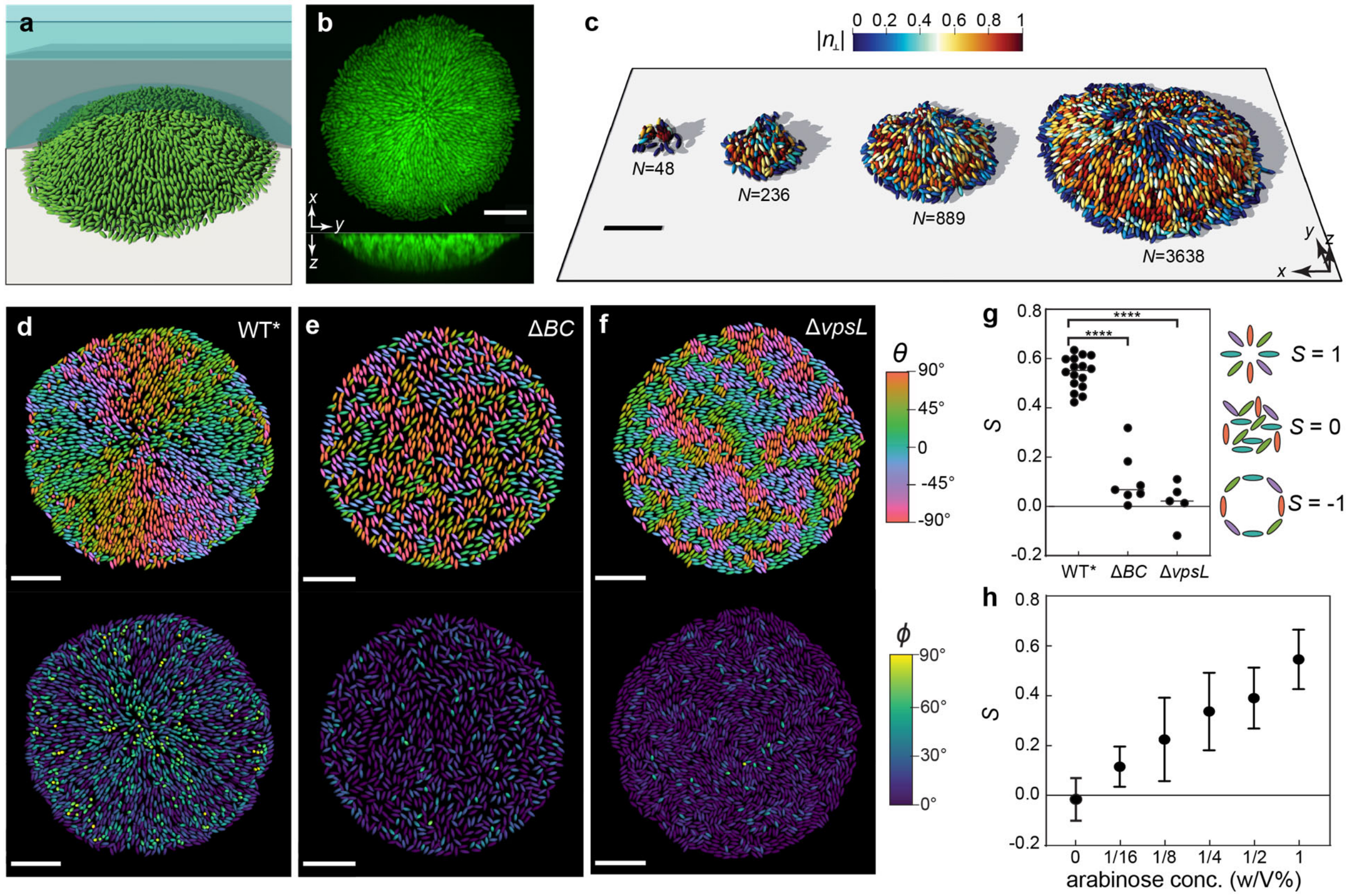

**Fig. 1| *V. cholerae* biofilms self-organize into aster patterns. a**, Schematic of the experimental setup, where *V. cholerae* biofilms (green) were grown on a glass surface covered by a hydrogel (blue shaded). **b**, Representative cross-sectional views of a WT* biofilm expressing mNeonGreen. **c**, Single-cell 3D reconstruction of biofilm structures over time with different numbers of cells $N$. **d-f**, Cell orientation color-coded according to each cell's angle in the basal plane $\theta$ (*Top*) or the angle it makes with the substrate $\phi$ (*Bottom*), in a biofilm that produces both exopolysaccharides and surface adhesion proteins (WT*; **d**), in a biofilm that only produces exopolysaccharides (Δ*BC;* **e**), and in a bacterial colony with neither exopolysaccharides nor surface adhesion (Δ*vpsL;* **f**). Scale bars, 10 µm. **g**, Radial order parameter $S$ quantifying the degree to which cells conform to an aster pattern in the three strains. Data was subjected to ANOVA for comparison of means. ****denotes P<0.0001. **h**, $S$ in biofilms in which the expression of *rbmC* is controlled by an arabinose inducible promotor. Error bars correspond to standard deviation.

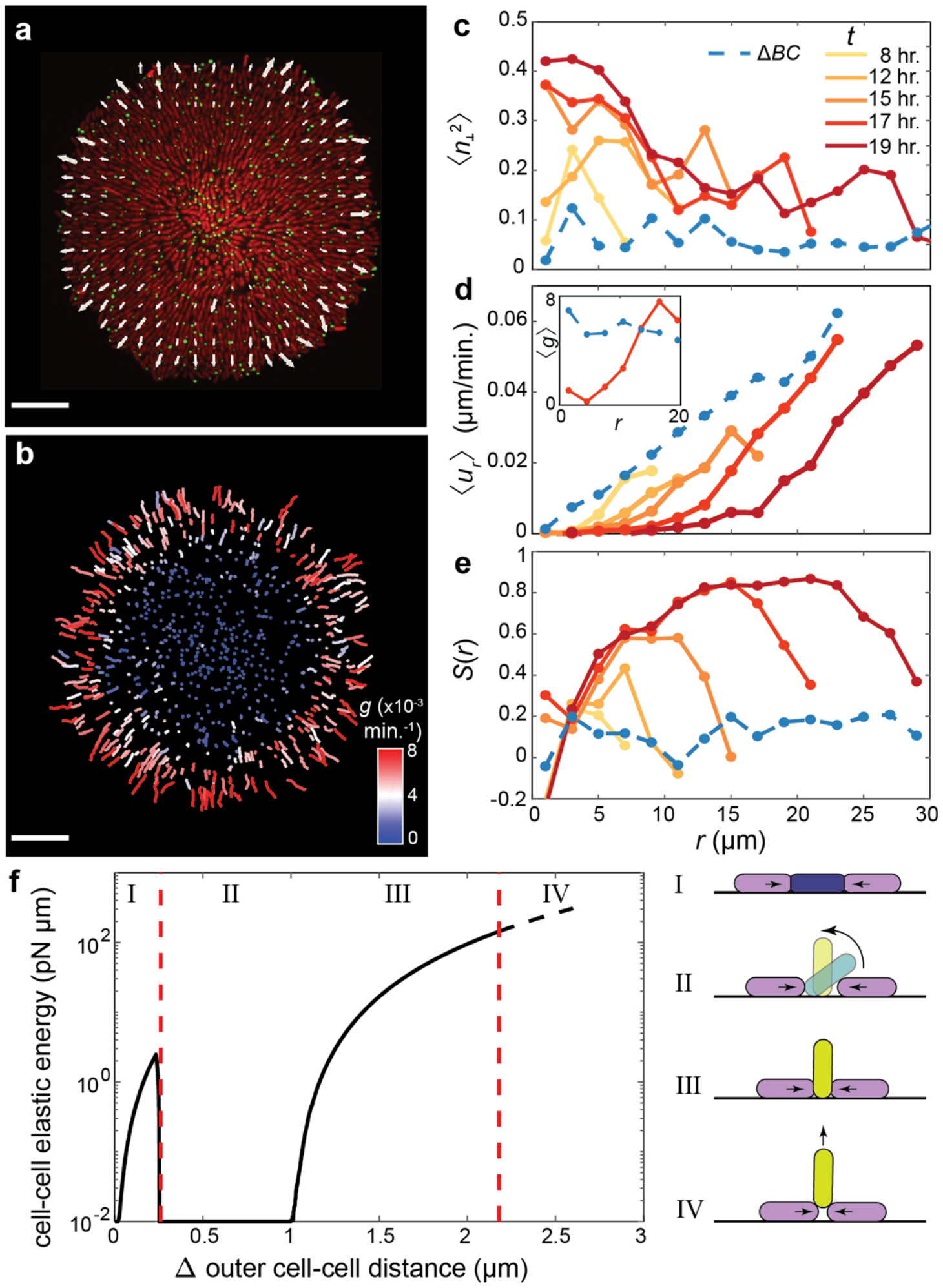

**Fig. 2| Growth-induced cellular flow and surface anchoring jointly lead to aster formation in biofilms. a**, Raw image of the basal layer of a biofilm consisting of cells constitutively expressing mScarlet-I cytosolically and mNeonGreen-labelled puncta. Overlain is the velocity field measured from puncta trajectories. **b**, Puncta trajectories colored by the apparent in-plane growth rate $g$. The apparent in-plane growth rate is calculated as $g(r) = (\partial_r r u_r)/r$ in a neighborhood around each cell. Scale bars, 10 μm. **c-e**, Azimuthally averaged degree of verticalization $\langle n_\perp^2 \rangle$ (**c**), radial velocity $\langle u_r \rangle$ (Inset: apparent in-plane growth rate $\langle g \rangle \times 10^{-3}\text{min.}^{-1}$) (**d**), and radial order parameter $S$ (**e**), as a function of distance $r$ from the center, in the basal layer. The dashed blue lines denote results from the nonadherent mutant. **f**, Results of a

reduced problem showing the strain energy due to cell-to-cell contacts in a cell as it is squeezed by two neighbors (black line). The dashed red lines denote the results from stability analyses (Supplementary Information Section 3). Upon increasing compression, the central cell evolves through four phases, which are given schematically.

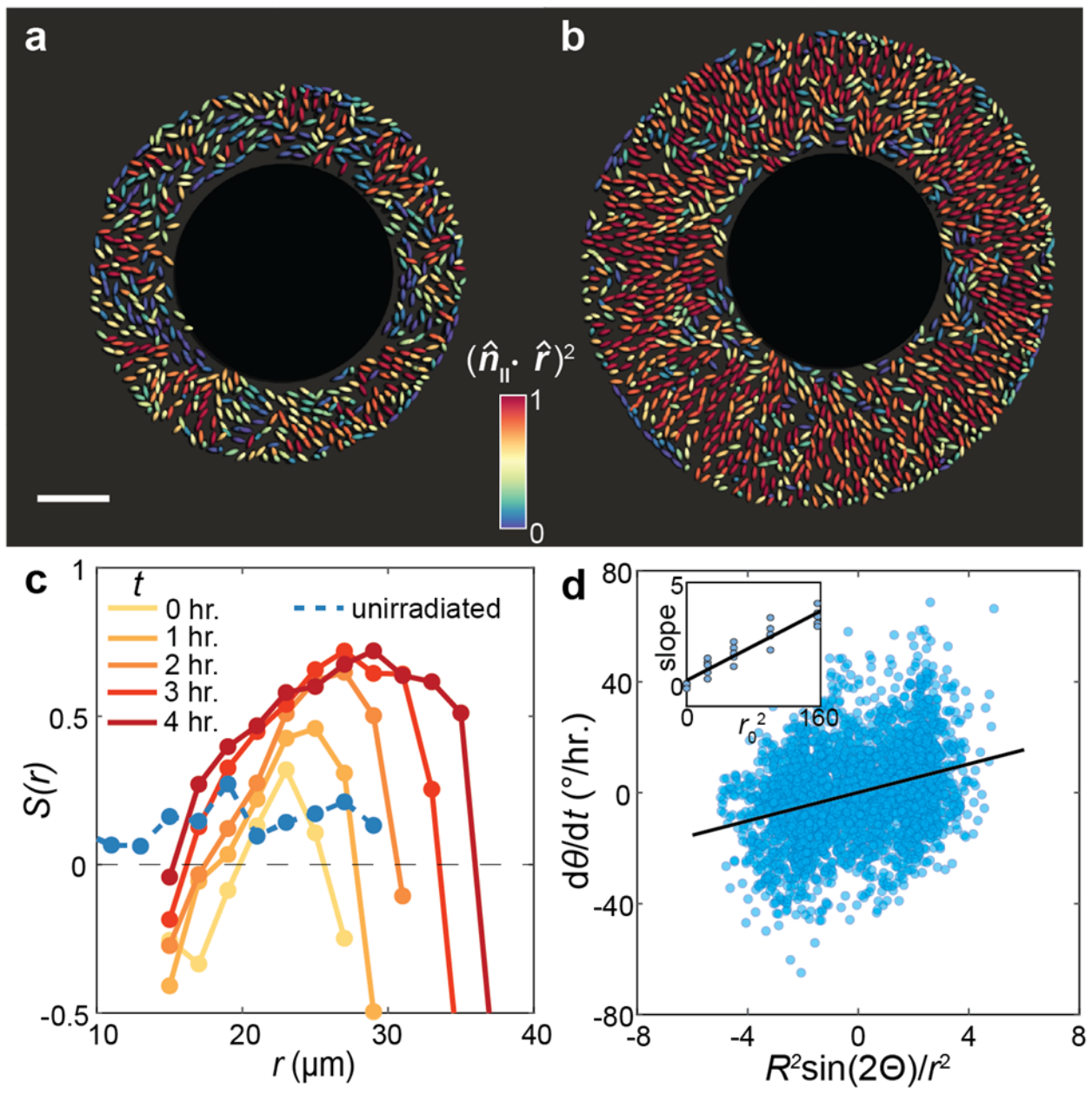

**Fig. 3| Cell organization can be manipulated by controlling spatial growth patterns. a**, **b**, A nonadherent biofilm grown for 17 hours was irradiated using 405 nm laser to induce cell death in a circle of radius 15 μm at the center. Colors denote the degree of radial alignment of individual cells $(\hat{\boldsymbol{n}}_{\parallel} \cdot \hat{\boldsymbol{r}})^2$. **c**, $S(r)$ in the irradiated biofilm (colored according to time) and the unirradiated control (blue). **d**, Angular velocity of individual cells from ABSs with a growth void plotted against the predicted nondimensionalized driving force. Inset: Fitted slope from **d** for different growth void sizes $r_0$ (μm$^2$).

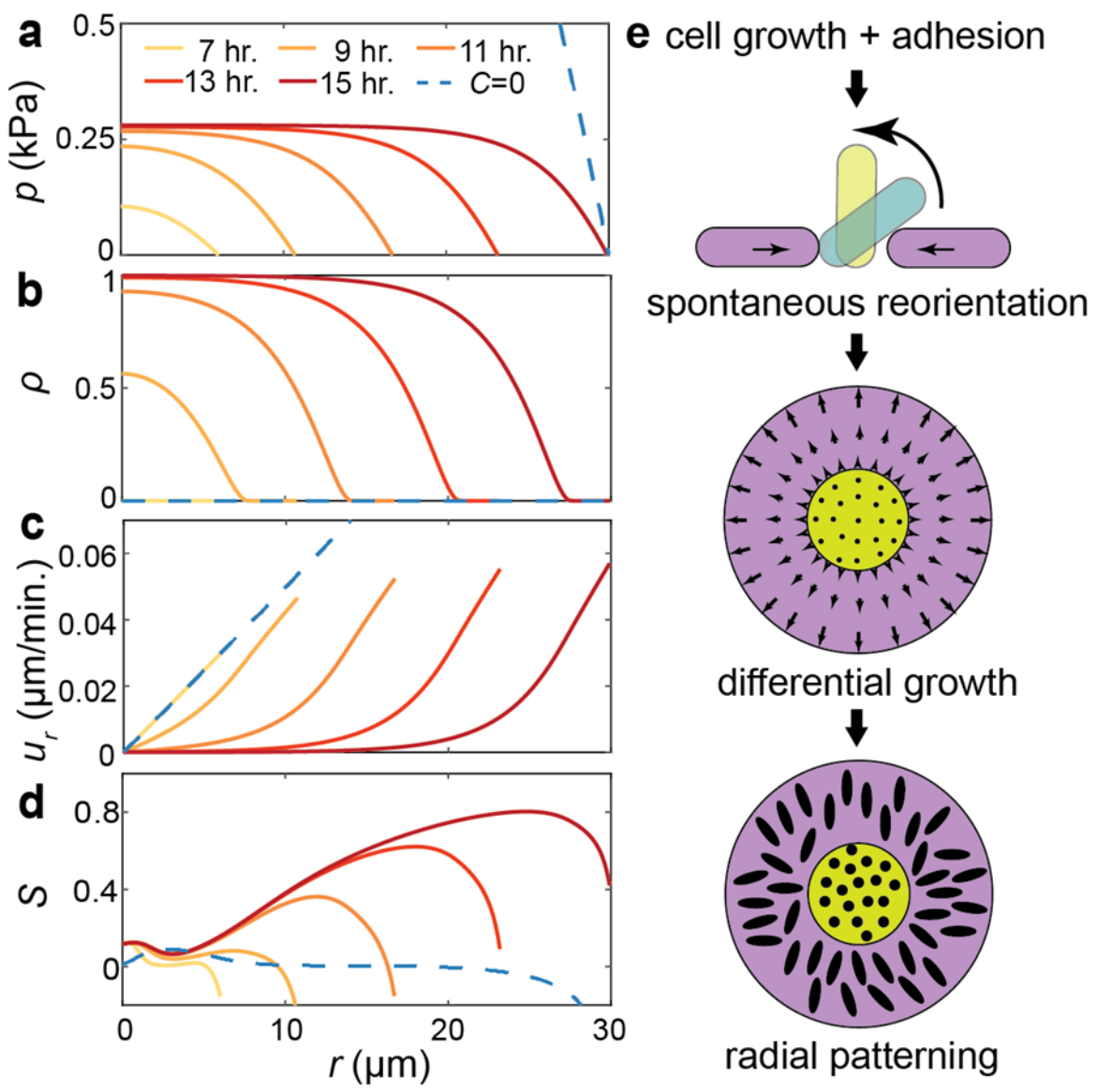

**Fig. 4| A two-phase active nematic model predicts spontaneous generation of differential proliferation and macroscopic cell organization. a-d**, Numerical solution of the model consisting of a population of horizontal and vertical cells. The biofilm was initiated with no vertical cells and random in-plane orientations. Evolution of pressure $p$ (**a**), fraction of vertical cells $\rho$ (**b**), in plane radial velocity $u_r$ (**c**), and radial order parameter $S$ (**d**). Curves are colored according to time. Results for a biofilm that cannot sustain verticalized cells ($C = 0$) are shown in blue. **e**, Schematic representation of the biofilm self-patterning process.

Supplementary Information for

# Biofilm self-patterning: mechanical forces drive a reorientation cascade

Japinder Nijjer, Changhao Li, Qiuting Zhang, Haoran Lu, Sulin Zhang*, Jing Yan*

*Corresponding authors. Email: suz10@psu.edu or jing.yan@yale.edu

## I. Cell segmentation algorithm

After the deconvolved 3D image (Extended Data Fig. 2a, b) was binarized (Extended Data Fig. 2c), a number of different connected components (CCs), each containing multiple cells, were identified. The key difficulty in segmentation lies in the inferior resolution in the *z*-direction in confocal images as well as imperfect rejection of out-of-focus light, which makes the use of a single global threshold inadequate. Therefore, we developed an adaptive local thresholding algorithm for single-cell biofilm image analysis. Specifically, for each CC, we gradually increased the intensity threshold locally and therefore delete voxels whose intensities are below the threshold. After each local binarization step, we checked whether new CCs were generated. For each newly generated CC, if its volume was below a defined volume threshold, we extracted this CC from the image and considered this CC as the core of one individual cell. Information on this individual cell was stored and no further action was performed on this cell until the image-restoration step. If the volume of the CC was still above the volume threshold, we locally increased the threshold until the volume fell below the threshold. After the adaptive local thresholding algorithm, we obtained the cores of all individual cells; however, these cores were smaller than the original cells; they did not represent accurately the location, orientation, and size of the original cells. Therefore, we added a volume restoration step in which the deleted voxels in the segmentation step were merged with the nearest core (Extended Data Fig. 2d, e). By performing the virtual shrinking-expansion step, we maintained the accuracy in both segmentation and in measuring the shape of each bacterium.

## II. Details of the agent-based simulations (ABSs)

*Single cell model*

We model the space occupied by a single cell and its surrounding extracellular matrix as a spherocylinder with length $L$, radius $R$ and therefore volume $V = \frac{4}{3}\pi R^3 + \pi R^2 L$. We assume the cell only grows in length over time while the radius remains the same, where the volume growth obeys the exponential law $\frac{\mathrm{d}V}{\mathrm{d}t} = \gamma V$, with growth rate $\gamma$. We introduce noise during cell growth and division, by assigning different cells with slightly different growth rates, taken from a normal distribution $\gamma \sim \mathrm{N}(\gamma_0, 0.2\gamma_0)$. In the ABSs, cell growth is implemented by increasing the length $L$ by the increment $\Delta L = \gamma\left(\frac{4}{3}R + L\right)\Delta t$, where $\Delta t$ is the timestep.

Cell division is modeled as the instantaneous replacement of a mother cell when it reaches length $L_{\text{max}}$ by two equal sized daughter cells. The slight volume loss after division is ignored to avoid contact overlapping which may cause unphysical reorientations. It follows that the initial length of a daughter cell is $L_0 = \frac{L_{\text{max}}}{2} - R$. Given the growth law, the doubling time can be calculated as $t_{\text{double}} = \frac{1}{\gamma}\log\left(\frac{10R+6L_0}{4R+3L_0}\right)$.

*Cell-cell repulsion*

We only consider the repulsive, elastic contact forces between cells, mediated by the soft exopolysaccharides. Linear elastic Hertzian contact theory is applied to quantify the repulsive contact force on cell $i$ by cell $j$, written as

$$\boldsymbol{F}_{\text{cell-cell},ij} = -\frac{5}{2}E_0 R^{1/2}\delta_{ij}^{3/2}\hat{\mathbf{e}}_{ij} \qquad \textbf{(1)}$$

where $E_0$ is the effective cell stiffness, $\delta_{ij}$ is the overlapping distance, and $\hat{\mathbf{e}}_{ij}$ is the unit vector parallel to the distance vector $\boldsymbol{d}$. The distance vector is given by connecting two contact points characterizing the smallest distance between two cell centerlines, as shown in Extended Data Fig. 4. The overlapping distance $\delta_{ij}$ is then calculated by

$$\delta_{ij} = 2R - |\boldsymbol{d}|, \qquad \textbf{(2)}$$

and contact only occurs if $\delta_{ij} > 0$. Taking the center of mass as the reference point, the moment of contact force is

$$\boldsymbol{M}_{\text{cell-cell},ij} = -\frac{5}{2}s_r\hat{\boldsymbol{n}}_i \times E_0 R^{1/2}\delta_{ij}^{3/2}\hat{\mathbf{e}}_{ij} \qquad \textbf{(3)}$$

where $\hat{\boldsymbol{n}}_i$ is the cell director and $s_r$ is the parametric coordinate, along the center-line of the cell, of the contact point.

*Cell-to-substrate interactions*

The experimental substrate is modeled as an infinite, two-dimensional rigid plane located at $z = 0$. We again assume linear elastic Hertzian contact mechanics to quantify the repulsive contact between cells and the substrate. We also add an attractive force which depends on the contact area to account for the matrix protein mediated cell-to-surface adhesion.

To formulate the repulsive contact between the cells and the substrate, we apply a generalized Hertzian contact formula that smoothly accounts for the cell orientation dependent contact energy. In this case, the elastic deformation energy is given by $E_{\text{el},i} = E_0 R^{1/2}\delta_i^{5/2}$ and the equivalent penetration depth is given by

$$\delta_i^{5/2} = \int_{-L/2}^{L/2}\left[R^{1/2}\left|\hat{\boldsymbol{n}}_{\parallel,i}\right|^2\delta^2(s) + \frac{4}{3}\left(1 - \left|\hat{\boldsymbol{n}}_{\parallel,i}\right|^2\right)\delta^{3/2}(s)\right]\mathrm{d}s \qquad \textbf{(4)}$$

where $\hat{\boldsymbol{n}}_{\parallel,i}$ is the normalized projection of the $i$th cell director onto the substrate. The overlap function $\delta(s)$ denotes the overlapping distance between the cell and the substrate at the local cell-body coordinate $-L/2 \leq s \leq L/2$. Then, the net force $\boldsymbol{F}_{\text{el},i}$ and moment $\boldsymbol{M}_{\text{el},i}$ from the cell-substrate elastic repulsion can be given by

$$\boldsymbol{F}_{\mathrm{el},i} = 2E_0R^{1/2}\int_{L/2}^{-L/2} \hat{\boldsymbol{z}}\cdot\left[R^{1/2}\left|\hat{\boldsymbol{n}}_{\parallel,i}\right|^2\delta(s) + \left(1-\left|\hat{\boldsymbol{n}}_{\parallel,i}\right|^2\right)\delta^{1/2}(s)\right]\mathrm{d}s \tag{5}$$

$$\boldsymbol{M}_{\mathrm{el},i} = 2E_0R^{1/2}\int_{L/2}^{-L/2} [s\hat{\boldsymbol{n}}_i\times\hat{\boldsymbol{z}}]\left[R^{1/2}\left|\hat{\boldsymbol{n}}_{\parallel,i}\right|^2\delta(s) + \left(1-\left|\hat{\boldsymbol{n}}_{\parallel,i}\right|^2\right)\delta^{1/2}(s)\right]\mathrm{d}s \tag{6}$$

where $\hat{\boldsymbol{z}}$ is the unit vector perpendicular to the substrate.

We take the adhesion energy between the cells and the substrate to be of the form $E_{\mathrm{ad},i} = -\Sigma_0 A_i$, where $\Sigma_0$ is the adhesion energy density and $A_i$ is the equivalent contact area. The equivalent contact area is given by

$$A_i = \int_{-L/2}^{L/2} a(s)\mathrm{d}s = \int_{-L/2}^{L/2}\left[R^{1/2}\left|\hat{\boldsymbol{n}}_{\parallel,i}\right|^2\delta^{1/2}(s) + \pi R\left(1-\left|\hat{\boldsymbol{n}}_{\parallel,i}\right|^2\right)\mathrm{H}\big(\delta(s)\big)\right]\mathrm{d}s \tag{7}$$

where $\mathrm{H}(\cdot)$ is the Heaviside step function. The net adhesive force $\boldsymbol{F}_{\mathrm{ad},i}$ and moment $\boldsymbol{M}_{\mathrm{ad},i}$ are:

$$\boldsymbol{F}_{\mathrm{ad},i} = -\Sigma_0\int_{-L/2}^{L/2}\hat{\boldsymbol{z}}\left[\frac{1}{2}R^{1/2}\left|\hat{\boldsymbol{n}}_{\parallel,i}\right|^2\delta^{-1/2}(s)\right]\mathrm{d}s - \hat{\boldsymbol{z}}\Sigma_0\pi R\left(1-\left|\hat{\boldsymbol{n}}_{\parallel,i}\right|^2\right) \tag{8}$$

$$\begin{aligned}\boldsymbol{M}_{\mathrm{ad},i} = -\Sigma_0\int_{-L/2}^{L/2}[s\hat{\boldsymbol{n}}_i\times\hat{\boldsymbol{z}}]\left[\frac{1}{2}R^{1/2}\left|\hat{\boldsymbol{n}}_{\parallel,i}\right|^2\delta^{-1/2}(s)\right]\mathrm{d}s \\ - [s_0\hat{\boldsymbol{n}}\times\hat{\boldsymbol{z}}]\Sigma_0\pi R\left(1-\left|\hat{\boldsymbol{n}}_{\parallel,i}\right|^2\right)\end{aligned} \tag{9}$$

where $s_0$ denotes the cell-body coordinate such that $\delta(s_0) = 0$.

*Viscosity*

We consider two sources of viscosity: a bulk viscous force due to the extracellular matrix environment and a surface viscous force due to the substrate. The environmental viscous force and moment are given by Stoke's law,

$$\boldsymbol{F}_{\mathrm{stokes},i} = -\eta_0\boldsymbol{u}_i \tag{10}$$

$$\boldsymbol{M}_{\mathrm{stokes},i} = -\eta_0\int_{L/2}^{L/2} s\hat{\boldsymbol{n}}_i\times(\boldsymbol{\omega}_i\times s\hat{\boldsymbol{n}}_i)\mathrm{d}s = -\frac{\eta_0}{12}\boldsymbol{\omega}_i L^3 \tag{11}$$

where $\eta_0$ is the environmental viscosity, $\boldsymbol{u}$ is the velocity of the center of mass, and $\boldsymbol{\omega}$ is the angular velocity. The substrate viscous force and moment are taken to be of the form

$$\boldsymbol{F}_{\mathrm{surface},i} = -\int_{-L/2}^{L/2}\frac{\eta_1 a(s)}{R}[\boldsymbol{u}_i(s) - (\boldsymbol{u}_i(s)\cdot\hat{\boldsymbol{z}})\hat{\boldsymbol{z}}]\mathrm{d}s \tag{12}$$

$$\boldsymbol{M}_{\mathrm{surface},i} = -\int_{-L/2}^{L/2}\frac{\eta_1 a(s)}{R}s\hat{\boldsymbol{n}}_i\times[\boldsymbol{u}_i(s) - (\boldsymbol{u}_i(s)\cdot\hat{\mathbf{z}})\hat{\boldsymbol{z}}]\mathrm{d}s \tag{13}$$

where $\eta_1$ is the viscous coefficient along the substrate.

*Equations of motion*

The equations of motion for each cell are given by Newton's rigid body dynamics:

$$\begin{pmatrix} \boldsymbol{F}_{\mathrm{net},i} \\ \boldsymbol{M}_{\mathrm{net},i} \end{pmatrix} = \begin{bmatrix} m & \mathbf{0} \\ \mathbf{0} & \mathbf{I}_i \end{bmatrix} \begin{pmatrix} \dot{\boldsymbol{u}}_i \\ \dot{\boldsymbol{\omega}}_i \end{pmatrix} + \begin{pmatrix} \mathbf{0} \\ \boldsymbol{\omega}_i \times \mathbf{I}_i \boldsymbol{\omega}_i \end{pmatrix} \quad \textbf{(14)}$$

where $\boldsymbol{F}_{\mathrm{net},i}$ and $\boldsymbol{M}_{\mathrm{net},i}$ are the total force and moment vector, and $\mathbf{I}$ is the moment of inertia. All of the variables are expressed in the body-fixed coordinate system, then transformed into the global coordinate system. We add a small random noise to the net force and moment vectors of the cells at every timestep ($10^{-7}E_0R^2$ for forces and $10^{-7}E_0R^3$ for moments).

*Choice of parameters for cellular dynamics*

The list of physical constants, which were used to successfully capture the verticalization instability in biofilm-dwelling cells in prior work[1], are used here and given in Table S2.

*Geometry of coarse-grained gel system*

We model the agarose gel using a coarse-grained particle-based approach. We treat the gel as a collection of spherical particles, each with radius $R_{\mathrm{gel}}$. We treat the interactions of the spheres using a spring network model to recapitulate the elastic behavior of the hydrogel. The pairwise interaction energy between the gel particles is $E_{\mathrm{gel},2} = \Sigma_{ij} \frac{k_r}{2} \left(\xi_{ij} - \xi_0\right)^2$, where $\xi_{ij}$ is the distance between particle $i$ and $j$, $\xi_0$ is the equilibrium distance, and $k_r$ is the spring constant. Furthermore, to impart a shear modulus to the system, we also include a three-body interaction energy where $E_{\mathrm{gel},3} = \Sigma_{ijk} \frac{k_\zeta}{2} \left(\zeta_{ijk} - \zeta_0\right)^2$, where $\zeta_{ijk}$ is the bond angle formed by particle $i$, $j$, and $k$.

*Interactions between gel particles and cells*

We again apply linear elastic Hertzian contact theory to describe the repulsive interaction between the coarse-grained gel particles and the cells. Similar to the contact between cells, the elastic contact energy can be written as $E_{\mathrm{gel-cell},ij} = E_1 \left(\frac{1}{R} + \frac{1}{R_{\mathrm{gel}}}\right)^{-1/2} \delta_{ij}^{5/2}$, where $E_1$ is the contact stiffness between gel and cell, and $\delta_{ij}$ is the overlapping distance between the $i$th gel particle and $j$th cell. The contact stiffness $E_1$ can be given by the relation $\frac{1}{E_1} = \frac{1-\mu_{\mathrm{gel}}^2}{Y_{\mathrm{gel}}} + \frac{1-\mu_{\mathrm{cell}}^2}{Y_{\mathrm{cell}}}$, where $Y$ and $\mu$ are Young's modulus and Poisson's ratio, respectively.

*Interactions between the gel particles and the substrate*

There exist two types of interactions between the coarse-grained gel particles and the substrate. First, the gel particles make elastic contact with the substrate, again described by Hertzian contact theory for the contact between a sphere and a flat surface, where the elastic contact energy is given by $E_{\mathrm{gel-surface},i} = E_1 R_{\mathrm{gel}}^{1/2} \delta_i^{5/2}$. Second, we introduce adhesion between the gel particles and substrate which provides an energy barrier to delamination; we take the energy to be of the form $E_{\mathrm{ad,gel},i} = -\Sigma_1 A_{\mathrm{gel},i}$, where $\Sigma_1$ is the adhesion energy density. The equivalent contact area is given by $A_{\mathrm{gel},i} = \pi R_{\mathrm{gel}} \delta_i$, where $\delta_i$ is the overlap between the gel particle and the substrate.

*Equation of motion*

We only consider the three translational degrees of freedom for gel particles, neglecting the rotational degrees of freedom. Therefore, the equations of motion are given by Newton's second law $\mathbf{F}_{\text{tot},i} = m\boldsymbol{a}_i$, where $\boldsymbol{F}_{\text{tot},i}$ is the net force and $\boldsymbol{a}_i$ is the acceleration. To prepare the initial amorphous stress-free geometry, we begin with a body-centered cubic crystalline geometry with lattice parameter $a$, where $a = 1.3R_{\text{gel}}$. Subsequently, we assigned the system with an initial temperature of $300K$ and annealed it until it reached a final configuration that is amorphous and stress-free.

*Choice of parameters*

The spring constant $k_r$ and equilibrium length $\xi_0$: The Young's modulus of the coarse-grained gel system is given by $Y = \frac{1}{V}\frac{\partial^2(E_{\text{gel},2}+E_{\text{gel},3})}{\partial\epsilon^2} \simeq \frac{k_r}{2\xi_0}^2$, under the condition $k_\zeta \ll k_r$. Generally, a smaller $\xi_0$ leads to a denser gel system and better approximation to a continuum solid. Here we choose $\xi_0 = 0.6$ μm as a result of a trade-off between simulation quality and computational cost, as the simulation time is proportional to $\frac{1}{{\xi_0}^3}$. This $\xi_0$ value leads to $k_r = 6 \times 10^{-3}\text{Nm}^{-1}$, corresponding to the experimentally measured $Y = 5kPa$.

Radius of coarse-grained gel particle $R_{\text{gel}}$: In order to mimic the continuum constraints posed by the hydrogel in the experiment, the coarse-grained system should not have significant defects larger than the volume of a single cell, which means that $R_{\text{gel}}$ should be larger than $\xi_0$. On the other hand, $R_{\text{gel}}$ cannot be significantly larger than the cell radius $R$, as this will introduce unphysical contacts at the biofilm-gel interface. Taking both requirements into consideration, we choose $R_{\text{gel}} = 1.0$ μm, which is nearly double the equilibrium distance $\xi_0 = 0.6$ μm and we keep $R_{\text{gel}} \approx R$.

Cell-gel contact modulus $E_1$: Hertzian contact mechanics gives the equivalent contact modulus between two elastic bodies by $\frac{1}{E_1} = \frac{1-\mu_{cell}^2}{Y_{cell}} + \frac{1-\mu_{gel}^2}{Y_{gel}}$, where $Y$ is the Young's modulus and $\mu$ is Possion's ratio of the two contacting elastic bodies. Given the bulk rheology measurement $Y_{\text{cell}} = 500$ Pa, $\mu_{\text{cell}} = \mu_{\text{gel}} = 0.49$, and $Y_{\text{gel}} = 5$ kPa[3,4], it follows $E_1 = 600$ Pa. In the simulations, we choose $E_1 = 1500$ Pa, somewhat larger than $E_0$, to avoid the unphysical situation in which cells penetrate into the gel during growth.

The three-body interaction parameter $k_\zeta$ and equilibrium angle $\zeta_0$: To make the amorphous coarse-grained gel system stable, we choose $\zeta_0 = 120°$ to force the system to deviate from its original cubic configurations. We select $k_\zeta$ such that we attain incompressible solid behavior in the gel $G = Y/3$ , where $G$ is the shear modulus and $Y$ is Young's modulus.

These parameters are listed in Table S3.

## III. Single-cell surface adhesion stability analysis

Previous studies have described how cells in 2D bacterial colonies can be peeled from the substrate through a buckling-like instability[5–8]; however, the reason why biofilm-dwelling cells self-organize into a verticalized core, that is, remain attached but oriented away from the substrate, remains to be shown. In this section, we consider the stability of surface-adhered cells in the presence of compression from neighboring cells to determine why surface adhesion facilitates stably surface-anchored, verticalized cells. Specifically, we consider a minimal model consisting of three spherocylinders following the same physics described in the agent-based modelling section above; however, we limit our analysis below to cells confined to the $xz$ plane, without any external gel confinement, and only experiencing the effects of the bulk viscosity for analytical tractability. We assume that the cells are not growing and instead mimic the effects of growth-induced compression by bringing the outer cells closer together. All three cells initially start out parallel to the substrate in close contact with each other, and the two outer cells are brought closer to each other to squeeze the central cell (see the schematic in Figure 2f). We find that the middle cell evolves through four phases upon increasing compression: **I.** the cell remains horizontal and the elastic potential energy in the cell increases; **II.** the middle cell rotates from horizontal to vertical and all potential energy is consequently released; **III.** the middle cell remains vertical, and the elastic potential energy increases again when the two outer cells touch the center cell; and **IV.** the middle cell becomes unstable and is ejected from the substrate. The transition from I to II corresponds to a "verticalization" instability[1], from II to III corresponds to a trivial geometric transition, and from III to IV corresponds to a "pinch-off" instability. Note that the pinch-off instability is locally (marginally) stable, but under sufficiently large $z$-perturbations, the middle cell gets ejected from the substrate. Experimentally, we expect that many factors could contribute to the perturbations including but not limited to: deviation of the cell shape from a perfect spherocylinder, fluctuating protein bonds, and variations in the adhesion protein concentration, etc. Importantly, we find that for the parameters listed in Table S2, the critical energy corresponding to the verticalization instability is much smaller than that of the pinch-off instability, leading to preferential verticalization of horizontal cells rather than pinch-off of verticalized cells (Fig. 2f). In what follows below, we consider these two instabilities in more detail paying specific attention to the role of the cell-to-surface adhesion.

The equations of motion for the position $\boldsymbol{r_i} = (x_i, z_i)$, and director $\hat{\boldsymbol{n}}_i = (n_x, n_z)$ of cell $i$ written using Lagrangian mechanics, are:

$$\eta_0 l \dot{\boldsymbol{r}}_i = -\frac{\partial \sum E_i}{\partial \boldsymbol{r_i}},$$
$$\frac{\eta_0 l^3 \dot{\hat{\boldsymbol{n}}}_i}{12} = -\frac{\partial \sum E_i}{\partial \hat{\boldsymbol{n}}_i} + \lambda \frac{\partial F_i}{\partial \hat{\boldsymbol{n}}_i}, \qquad \textbf{(15A,B)}$$

where $\lambda$ is a Lagrange multiplier corresponding to the constraint $F_i = |\hat{\boldsymbol{n}}_i| - 1 = 0$. Here $\sum E_i$ is the total potential energy of cell $i$ and includes cell-to-cell repulsion, as well as cell-to-surface adhesion and repulsion terms.

In phase I, in the absence of any compression from neighboring cells, the balance of surface adhesion and repulsion leads to an equilibrium cell configuration $\hat{\boldsymbol{n}} = (1,0)$ at an apparent height

$$z_{0,h} = R\left(1 - \left(\frac{\tilde{A}}{4}\right)^{\frac{2}{3}}\right), \qquad \textbf{(16)}$$

where $\tilde{A} = \Sigma_0 / R E_0$ is the dimensionless adhesion coefficient. In the presence of compression, we consider perturbations away from this stable point. Taking $x$ as the distance between the centers of the two outer cells (we assume that the outer cells are at the same equilibrium height), the elastic energy in the cell of interest due to cell-to-cell contacts is $E_{\text{cc}} = 2E_0 R^{1/2}(2R - d)^{5/2}$ where the shortest distance between the cells is $d = L + 2R - x/2$. Combining with the cell-surface adhesion and cell-surface repulsion terms, we find the evolution equation for small $n_z$ is given by

$$\dot{n}_z = \Omega(x) n_z, \tag{17}$$

where the sign of $\Omega$ determines the stability of the fixed-point. Extended Data Figure 6a shows plots of $\Omega$ for different dimensionless adhesion coefficients. For small values of $\tilde{A}$, which are most physically relevant, the zeros of $\Omega$ are only weakly dependent on $\tilde{A}$. This analysis suggests that the verticalization instability only weakly depends on adhesion, that is, the energy barrier to verticalization is nearly constant.

In phase III, again assuming no compression from neighboring cells, the balance of surface adhesion and repulsion leads to an equilibrium verticalized configuration $\hat{\boldsymbol{n}} = (0,1)$ with height

$$z_{0,v} - \frac{L}{2} = R\left(1 - \left(\frac{3\pi\tilde{A}}{4}\right)^{2/3}\right). \tag{18}$$

Note that the effective penetration depth (the degree to which the soft cell is flattened against the substrate) is larger for verticalized cells than for horizontal cells. Again, we consider compression due to two outer horizontal cells (at a height $z_{0,h}$). Here we look for finite-sized perturbations away from the fixed point; the evolution equation of the height of the verticalized cell is given by

$$\dot{z} = w(x, z). \tag{19}$$

We look for points $(x, z)$ such that $w(x, z) > 0$. We only consider points $0 \le \frac{L}{2} + R - z < R - z_{0,h}$ as points $z - \frac{L}{2} < z_{0,h}$ results in no net vertical force such that the central cell remains stably attached to the substrate. Specifically, we focus on points very close to $z = \mathrm{L}/2 + R$ to determine the critical energy required to produce marginally unstable cells. Extended Data Figure 6b shows plots of $w(x, L/2 + 0.999R)$ for different adhesion coefficients. Contrary to the verticalization transition, in this case we find a strong dependence of the zeros of $w$ on $\tilde{A}$; therefore, the energy barrier to pinch-off depends strongly on cell-to-surface adhesion.

We combine these two stability analyses into a single plot of the energy landscape experienced by the center cell for different adhesion energies (Extended Data Figure 6c). We see that for small adhesion energies, the difference in the two energy barriers is small, whereas for large adhesion energies, the pinch-off instability requires orders of magnitude more energy than the verticalization instability. This energy difference between the two instabilities leads to more stably adhered verticalized cells. This analysis explains the observed positive correlation between the adhesion energy and the fraction of stably verticalized cells (Extended Data Fig. 3, 4): adhesion energy affects the stability of cells with respect to the pinch-off instability but much less so to the verticalization instability.

## IV. Continuum model for growth-induced macroscopic cell ordering

In this section, we present a minimal coarse-grained description to explain the growth and self-organization observed in the basal layer of *V. cholerae* biofilms. We first consider the simpler case with one population of cells that either generate in-plane growth or not, before considering the complete two-phase model which accounts for pressure-dependent cell verticalization and out-of-plane growth.

*Theoretical framework*

We limit our consideration to the basal plane of the biofilm and assume it is a growing, quasi-2D system with *mesoscopic* nematic order tensor $\boldsymbol{Q} = 2\langle\hat{\boldsymbol{n}}\otimes\hat{\boldsymbol{n}} - \boldsymbol{I}/2\rangle$[9]. Based on its symmetries, in 2D, $\boldsymbol{Q}$ can also be written as

$$\boldsymbol{Q} = q\begin{bmatrix}\cos(2\theta) & \sin(2\theta)\\ \sin(2\theta) & -\cos(2\theta)\end{bmatrix} \quad \textbf{(20)}$$

where $\theta$ is the angle of the average head-less director and $q$ is the scalar order parameter, which quantifies the degree to which the cells are locally ordered. In the analysis below we will assume that this scalar order parameter is constant.

We assume the bacterial cells grow with a uniform growth rate $\gamma$ but that the growth can be out-of-plane, resulting in a spatially varying in-plane growth rate $g(\boldsymbol{r})$. Taking the cell density as $c$, and the coarse-grained growth-induced velocity field as $\boldsymbol{u}$, mass conservation requires

$$\partial_t c + \nabla\cdot(c\boldsymbol{u}) = gc. \quad \textbf{(21)}$$

The density is nearly uniform in the experiment, that is, $c(\boldsymbol{r}) = c_0 + \epsilon c_1(\boldsymbol{r})$, where $\epsilon \ll 1$. To leading order, Eq. (21) becomes

$$\nabla\cdot\boldsymbol{u} = g. \quad \textbf{(22)}$$

The growth-induced velocity field is found by integrating the spatially dependent growth rate. In this analysis, we will neglect density fluctuations $c_1$, but it can be related to the fluctuation in local pressure due to varying local configurations.

The evolution of the nematic order tensor is described by the Beris-Edward equation[10]

$$(\partial_t + \boldsymbol{u}\cdot\nabla)\boldsymbol{Q} = \lambda\boldsymbol{E} - (\boldsymbol{\omega}\cdot\boldsymbol{Q} - \boldsymbol{Q}\cdot\boldsymbol{\omega}) + \Gamma\boldsymbol{H}, \quad \textbf{(23)}$$

where $\lambda$ is the flow-alignment parameter and $\lambda > 1$ for rod-shaped objects that tend to align in shear. Here $\boldsymbol{E} = \frac{1}{2}(\nabla\boldsymbol{u} + \nabla\boldsymbol{u}^T - (\nabla\cdot\boldsymbol{u})\boldsymbol{I})$ and $\boldsymbol{\omega} = \frac{1}{2}(\nabla\boldsymbol{u} - \nabla\boldsymbol{u}^T)$ are the traceless strain-rate and vorticity tensors and $\boldsymbol{H}$ is the molecular field. $\Gamma\boldsymbol{H}$ relaxes $\boldsymbol{Q}$ towards a bulk state with minimal spatial variation in the director. However, we do not expect $\Gamma\boldsymbol{H}$ to play an important role in biofilm ordering because biofilm-dwelling cells secrete exopolysaccharides which are soft and deformable and act as a "cushion" between cells. This can be seen in the biofilm from the $\Delta BC$ mutant (Fig. 1e) in which minimal local alignment was observed. Contrast this with the $\Delta vpsL$ mutant and other bacterial colonies[11–13] where there is significant local alignment (Fig. 1f) when no exopolysaccharides are present. Finally, force balance requires

$$\nabla\cdot\boldsymbol{\Pi} = \eta\boldsymbol{u}, \quad \textbf{(24)}$$

where $\boldsymbol{\Pi}$ is the stress tensor in the biofilm, whose divergence is balanced by surface friction. Here we have assumed that energy dissipation is dominated by surface friction rather than viscous dissipation inside the biofilm, corresponding to the so-called "dry" limit[14]. We take the frictional force density to be linearly proportional to the velocity and the friction coefficient $\eta$ to be isotropic and constant. In many active nematic systems, the active stress is anisotropic and dependent on $\boldsymbol{Q}$. For instance, in motile cell colonies, cells generate anisotropic active stresses[15–17] while in growing bacterial colonies a number of different constitutive relations have been proposed[11,12,18,19]. Although at the microscopic scale, these forces may be anisotropic, on the mesoscopic scale pertaining to our theory, they are expected to be isotropic on average. This leads to the assumption of an isotropic stress field which is balanced by surface friction,

$$\nabla p = \eta \boldsymbol{u}. \quad \textbf{(25)}$$

Pressure in this system arises due to compression from neighboring cells, mediated by the exopolysaccharides, which arises from cell proliferation.

To close the model, we impose the following set of boundary and initial conditions. By symmetry, we expect that the velocity goes to zero at the center of the biofilm,

$$\boldsymbol{u}(\boldsymbol{0}, t) = \boldsymbol{0}, \quad \textbf{(26)}$$

that is, the center of the biofilm does not move and

$$\frac{\partial \boldsymbol{Q}}{\partial r}(\boldsymbol{0}, t) = \boldsymbol{0}. \quad \textbf{(27)}$$

We also assume that the pressure accumulates towards the center of the biofilm and that at the outer edge of the biofilm the pressure is zero,

$$p(r = r_e, t) = 0, \quad \textbf{(28)}$$

where $r_e$ denotes the edge of the biofilm. The outer radius evolves through the kinematic condition

$$\frac{dr_e}{dt} = u_r(r_e). \quad \textbf{(29)}$$

Finally, we assume that the orientation of the biofilm is described by some initial order parameter tensor

$$\boldsymbol{Q}(r, \psi, t = 0) = \boldsymbol{Q_0}(r, \psi), \quad \textbf{(30)}$$

which can also be written as $\theta(r, \psi, t = 0) = \theta_0(r, \psi)$.

Equations 22, 23, 25 and boundary conditions and initial conditions Eq. 26-30 form the basis of the theoretical model. In the following subsections, we consider a few idealized cases to determine under which conditions biofilms self-organize into an aster pattern. Throughout, we will also make the simplifying assumption that all variables except for $\boldsymbol{Q}$ are axisymmetric and that there is no azimuthal flow, $\boldsymbol{u} = u_r \hat{r}$.

<u>Uniform growth limit</u>

If all growth is in-plane, then $g = \gamma$ is uniform and constant. Integrating Eq. 22 we find the velocity field is

$$u_r = \frac{\gamma r}{2}, \quad \textbf{(31)}$$

and the strain-rate and vorticity tensors are exactly zero, $\boldsymbol{E} = \boldsymbol{\omega} = \boldsymbol{0}$. Solving Eq. 23 yields

$$\boldsymbol{Q}(r, \psi, t) = \boldsymbol{Q}_0\left(re^{-\frac{\gamma t}{2}}, \psi\right). \quad \textbf{(32)}$$

*In this case,* $\boldsymbol{Q}$ *is simply stretched by growth, and there is no tendency for* $\boldsymbol{Q}$ *to rotate or align in any given direction.* An initially isotropic system will remain isotropic as it grows (Extended Data

Fig. 7a-c)[12,20]. This is what happens in the 2D expansion of bacterial colonies, which remain macroscopically isotropic as they grow. Although microscopic forces can align cells locally, they cannot introduce any long-range order. Note that in two dimensions axisymmetric growth is a special case where there is no net strain in the system: if instead the system were confined into a rectangular channel, where growth is unidirectional, the strain rate tensor would not be zero and therefore there would be net alignment along the direction of the channel[18,19,21,22].

*Alignment due to a growth void*

Now suppose there is a radius $r_0$ inside which the cells have verticalized and are angled out of plane. These cells do not contribute to the growth of the basal plane and so

$$g(r) = \begin{cases} 0 \text{ if } r < r_0 \\ \gamma \text{ if } r \geq r_0 \end{cases}. \quad \textbf{(33)}$$

Substituting into Eq. 22, the growth-induced velocity is

$$u_r = \begin{cases} 0 & \text{if } r < r_0 \\ \gamma(r - r_0^2/r)/2 & \text{if } r \geq r_0 \end{cases}. \quad \textbf{(34)}$$

The corresponding strain rate and vorticity tensors in polar coordinates ($\boldsymbol{E_p}, \boldsymbol{\omega_p}$) are

$$\begin{aligned} &\boldsymbol{E_p} = \boldsymbol{0}, && \boldsymbol{\omega_p} = \boldsymbol{0} \text{ if } r < r_0 \\ &\boldsymbol{E_p} = \begin{bmatrix} \left(\frac{\partial u_r}{\partial r}\right) - \frac{\gamma}{2} & 0 \\ 0 & \frac{u_r}{r} - \frac{\gamma}{2} \end{bmatrix}, && \boldsymbol{\omega_p} = \boldsymbol{0} \text{ if } r \geq r_0. \end{aligned} \quad \textbf{(35)}$$

Under uniform growth $\partial u_r/\partial r$ and $u_r/r$ are equal to $\gamma/2$ , leading to no net deviatoric strain-rate. In contrast, in the presence of a growth void, these two terms become unequal, leading to a net deviatoric strain-rate. This is because the radial gradient in velocity is no longer balanced by azimuthal stretching due to outward expansion. For a velocity of the form given in Eq. 34, the strain rate and vorticity tensors in polar coordinates are

$$\begin{aligned} &\boldsymbol{E_p} = \boldsymbol{0}, && \boldsymbol{\omega} = \boldsymbol{0} \text{ if } r < r_0 \\ &\boldsymbol{E_p} = \frac{\gamma r_0^2}{2r^2}\begin{bmatrix} 1 & 0 \\ 0 & -1 \end{bmatrix}, && \boldsymbol{\omega} = \boldsymbol{0} \text{ if } r \geq r_0. \end{aligned} \quad \textbf{(36)}$$

Next, the nematic order parameter $\boldsymbol{Q}$ can be transformed to a polar coordinate system, given by

$$\boldsymbol{Q_p} = \boldsymbol{R^T Q R} = q\begin{bmatrix} \cos(2(\theta - \psi)) & \sin(2(\theta - \psi)) \\ \sin(2(\theta - \psi)) & -\cos(2(\theta - \psi)) \end{bmatrix}, \quad \textbf{(37)}$$

where $\psi$ is the azimuthal angle and $\boldsymbol{R} = \begin{bmatrix} \cos\psi & -sin\,\psi \\ sin\,\psi & \cos\psi \end{bmatrix}$ is the coordinate transformation matrix from cartesian to polar coordinates.

Finally, given $\boldsymbol{E_p}$, $\boldsymbol{\omega_p}$ and $\boldsymbol{Q_p}$, the matrix form of Eq. (23) in polar coordinates is

$$q\left(\partial_t + u_r\frac{\partial}{\partial r}\right)\begin{bmatrix} \cos(2(\theta - \psi)) & \sin(2(\theta - \psi)) \\ \sin(2(\theta - \psi)) & -\cos(2(\theta - \psi)) \end{bmatrix} = \frac{\lambda\gamma r_0^2}{2r^2}\begin{bmatrix} 1 & 0 \\ 0 & -1 \end{bmatrix}, \quad \textbf{(38)}$$

for $r \geq r_0$, which yields two scalar equations

$$-\sin(2(\theta-\psi))\left(\frac{\partial\theta}{\partial t}+u_r\frac{\partial\theta}{\partial r}\right)=\frac{\lambda\gamma r_0^2}{4qr^2}, \qquad \textbf{(39A,B)}$$
$$\cos(2(\theta-\psi))\left(\frac{\partial\theta}{\partial t}+u_r\frac{\partial\theta}{\partial r}\right)=0.$$

Defining $\Theta=(\theta-\psi)$ and combining the two scalar equations above, we find that for $r \geq r_0$,

$$\partial_t\Theta+u_r\partial_r\Theta= -\frac{\lambda\gamma r_0^2}{4qr^2}\sin(2\Theta). \qquad \textbf{(40)}$$

By first ignoring the advective term, we see that $\partial_t\Theta \sim -\sin(2\Theta)$ which has stable fixed points at $\Theta = n\pi$ or $\theta=\psi+n\pi$, characteristic of an aster pattern. *In other words, the rightmost term, which arises due to gradients in the growth-induced velocity field, drives cells to reorient towards an aster pattern.* The same observation was made during inward growth in bacterial colonies where deviation of the velocity field from that of isotropic growth resulted in radial alignment of cells[20].

More generally, if we suppose that instead of a growth void, there is a general differential growth rate: $\gamma$ for $r < r_0(t)$ and $\gamma+\Delta\gamma$ for $r \geq r_0(t)$ then the evolution equation for $\Theta$ for $r > r_0$ is given by

$$\partial_t\Theta+u_r\partial_r\Theta= -\frac{\lambda\Delta\gamma r_0^2}{4qr^2}\sin(2\Theta). \qquad \textbf{(41)}$$

If the biofilm has a faster growing outer rim $\Delta\gamma > 0$, the bacteria will tend to organize into an aster pattern. If, on the other hand, the biofilm has a faster growing core, i.e. $\Delta\gamma < 0$, $\Theta$ will have stable fixed points at $\Theta = n\pi + \pi/2$, characteristic of a vortex pattern. This means that biofilms that exhibit excess growth in the core will be driven to form a vortex.

Returning to Eq. 40, we solve it by the method of characteristics, assuming some initial maximum biofilm radius $r_e$, which yields

$$\cot[\Theta(r,\psi)]=\cot[\Theta_0(r',\psi)]\exp\left[\frac{\lambda}{2q}\left(\gamma t+\log\left(\frac{r'^2}{r^2}\right)\right)\right], \qquad \textbf{(42)}$$

where $r' = \left((r^2-r_0^2)e^{-\gamma t}+r_0^2\right)^{1/2}$ and $\Theta_0=\theta_0-\psi$ corresponds to the angular representation $\boldsymbol{Q}_0$. Plots of Eq. 42 are given in Extended Data Figures 7d-f. Note that, although cells tend to align in an aster pattern, the aligning torque becomes weaker as cells are advected outwards. Therefore, $S$ does not necessarily reach a value of 1 (Extended Data Fig. 7g). Depending on the initial size of the growth void as well as the flow alignment parameter, different degrees of radial alignment are reached in the final state, with a larger $r_0/r_e$ leading to overall more radial alignment.

To be more general, now instead of a finite void, suppose there is an arbitrary velocity field $u_r(r,t)$ corresponding to an arbitrary growth rate $g(r,t)$. Following the same procedure as above, we find the evolution equation for $\Theta$ is

$$\partial_t\Theta+u_r\partial_r\Theta= -f(r,t)\sin(2\Theta), \qquad \textbf{(43)}$$

where $f=(\lambda r/4q)\partial_r(u_r/r)$ quantifies the aligning torque due to gradients in the flow field. Equation (43) thus describes a generalized time-evolution equation for the orientation field due to gradients in the flow velocity.

*Two-phase active nematic model for cell verticalization and alignment*

We expand the active nematic model in the previous subsection to include the verticalization dynamics in a two-phase model consisting of vertical and horizontal cells. Consider a quasi-2D layer of cells on a substrate where the cells are either parallel (horizontal) or tilted (vertical) with respect to the substrate and let $\rho$ denote the fraction ($0 \leq \rho \leq 1$) of vertical cells and $1 - \rho$ therefore denote the fraction of horizontal cells. Horizontal cells proliferate and generate their progeny in the basal layer of interest, while vertical cells do not contribute to the expansion to the basal layer. The growth-induced stress generated by the horizontal cells also cause them to verticalize with a pressure dependent rate $C(p)$. The evolution equation for the fraction of these two populations of cells is therefore given by:

$$\begin{gathered} \partial_t \rho + \nabla \cdot (\boldsymbol{u}\rho) = C(p)(1-\rho) \\ \partial_t (1-\rho) + \nabla \cdot \big(\boldsymbol{u}(1-\rho)\big) = \gamma(1-\rho) - C(p)(1-\rho). \end{gathered} \qquad \textbf{(44A,B)}$$

Given that cell verticalization requires a minimum threshold pressure $p_t$, we make the simple assumption that the conversion rate is a linear function of the excess pressure

$$C(p) = \beta \frac{p - p_t}{p_t} H(p - p_t), \qquad \textbf{(45)}$$

where $H$ denotes the Heaviside function. Combining Eq. 44A and 44B yields

$$\nabla \cdot \boldsymbol{u} = \gamma(1-\rho). \qquad \textbf{(46)}$$

We initiate the biofilm with some initial radius $r_{e,0}$ and assume all cells are horizontal to begin with, $\rho(r, t = 0) = 0$. Finally, we assume symmetry at the origin which requires $\frac{\partial \rho}{\partial r} = 0$ and that $\rho$ is axisymmetric. Equations 25, 43, 44, 45, 46 with the above mentioned initial and boundary conditions make up the full two-phase continuum model for cellular ordering driven by verticalization.

*Non-dimensionalization*

We non-dimensionalize Eq. 25, 43, 44, 45, 46 by the following dimensional scales: time $t_d = 1/\gamma$, pressure $p_d = p_t$, length $r_d = (p_t/\eta\gamma)^{1/2}$ and velocity $u_{r,d} = r_d\gamma$, which yields

$$\frac{\partial \tilde{p}}{\partial \tilde{r}} = \tilde{u}_r, \qquad \textbf{(47)}$$

$$\frac{\partial \Theta}{\partial \tilde{t}} + u_{\tilde{r}} \frac{\partial \Theta}{\partial \tilde{r}} = -\tilde{L}\tilde{r} \frac{\partial \tilde{u}_r/\tilde{r}}{\partial \tilde{r}} \sin(2\Theta), \qquad \textbf{(48)}$$

$$\frac{\partial \rho}{\partial \tilde{t}} + \frac{1}{r} \frac{\partial (\tilde{r}\tilde{u}_r \rho)}{\partial \tilde{r}} = \tilde{\beta}(1-\rho)(p-1)H(p-1), \qquad \textbf{(49)}$$

$$\frac{1}{r} \frac{\partial (\tilde{r}\tilde{u}_r)}{\partial \tilde{r}} = 1 - \rho. \qquad \textbf{(50)}$$

with boundary conditions $p(\tilde{r}_e) = 0$, $\tilde{u}_r(0) = 0$, $d_t\tilde{r}_e = \tilde{u}_r(\tilde{r}_e)$, $\partial_{\tilde{r}}\Theta|_{\tilde{r}=0} = 0$, and $\partial_{\tilde{r}}\rho|_{\tilde{r}=0} = 0$ and initial conditions $\Theta(\tilde{r}, \psi, 0) = \Theta_0(\tilde{r}, \psi)$ and $\rho(\tilde{r}, 0) = 0$. At the boundary, when solving for $\Theta$, we impose a sharp velocity gradient $\Delta\tilde{u}_r/\Delta\tilde{r} \approx -\tilde{u}_r(\tilde{r}_e)/(\tilde{l}_{\text{Cell}}/2)$ where $\tilde{l}_{\text{Cell}}$ is the average length of a cell, to account for the fact that across the boundary cells there is a sharp velocity gradient which tends to align cells tangentially to the boundary[7,12]. There are two key parameters that control the evolution of the system: the dimensionless flow alignment parameter $\tilde{L} = \lambda/4q$ and the dimensionless verticalization rate $\tilde{\beta} = \beta/\gamma$. The flow alignment parameter dictates how quickly the cells will align to a straining flow, while the verticalization rate dictates how quickly horizontal cells are converted to vertical cells. For simplicity, we assume that the only distinguishing feature of the nonadherent mutant is that $\beta = 0$; however, we would also expect that in the absence of adhesion, $\eta$ would also be smaller and lead to an underestimation of the characteristic length-scale.

*Choice of parameters*

We choose the following set of parameters when solving for the evolution of the system.

Growth rate $\gamma$: The number of bacteria in a three-dimensional biofilm was measured over time and the growth rate was found to be $\gamma \approx 0.6$ hr.$^{-1}$

Threshold verticalization pressure $p_t$: From the theoretical verticalization analysis, using experimentally calibrated parameters, it was found that the critical cell overlap distance at which point verticalization happens is $\delta{\sim}0.125$ µm. From this value, the critical cell-cell contact pressure can be calculated as $p_d = 4E\delta^{\frac{1}{2}}/(3\pi R^{*1/2})$ where $E$ is the contact modulus of the cells which we estimate to be $500$ Pa using bulk rheological measurements (shear modulus of a bulk WT* biofilm was measured to be $\sim 200$ Pa) and $R^* = R/2 = 0.4$ µm. This yields a threshold pressure of $p_t \sim 130$ Pa.

Friction coefficient $\eta$: Using the surface viscosity $\eta_1$ defined in Table S2, we estimate the friction coefficient as $\eta = \eta_1/A$ where $A$ is the surface footprint of the cell which we estimate to be $A \sim 3$ µm$^2$. The friction coefficient is therefore $\eta \sim 7 \times 10^4$ Pa s/µm$^2$.

Characteristic length scale $r_d$: The characteristic length-scale $r_d = (p_t/\eta\gamma)^{1/2}$ in the system corresponds to the characteristic size of biofilm when verticalization begins. For the parameters defined above, this length scale is $r_d \sim 3.4$ µm.

Mean cell length $l_{\text{Cell}}$: Taking the average of the largest and smallest cylindrical sections of the spherocylindrical cells (corresponding to right before and right after division) gives a length of $l_{\text{Cell}} = 1.8$ µm.

Initial radius of biofilm $r_{e,0}$: We initialize the biofilm to have an area equivalent to the footprint of a single cell. This gives a value $r_{e,0} \sim 1$ µm.

We solve the dimensionless Eq. 47-50 using a finite difference scheme with explicit time-stepping, and up-winding of advective terms. Example solutions are given in Extended Data Figure 10. We fit the two unknown dimensionless parameters to the experimental results and find that $\tilde{\beta} \approx 2.5$ and $\tilde{L} \approx 1.5$. These parameters compare favorably to what has been measured before. Namely, Beroz *et al.*[1] measured in ABS that in a biofilm growing without a confining agarose gel, $\beta \sim 2.5$ hr.$^{-1}$. Here we find a somewhat smaller value, $\beta \sim 1.5$ hr.$^{-1}$, likely owing to the fact that confinement due to the agarose gel tends to impart normal stresses that suppress verticalization. Nonetheless, we find that the results are only weakly dependent on $\beta$ for $\beta > \gamma$ (Fig. 4a-d; Extended Data Fig. 10d-k). Finally, we estimate $q$ by measuring the cell-cell orientation correlation for neighboring cells in the basal layer of the experimental biofilms and find $q \sim 0.5$ and so $\lambda \sim 3$. Measuring the intrinsic flow alignment parameter is difficult as it is dependent on many factors including the size, shape, aspect ratio and local nematic order[23], however, a value of 3 is in reasonable agreement with values of 0.3-2 reported for other growing and non-growing bacterial systems[11,18,24].

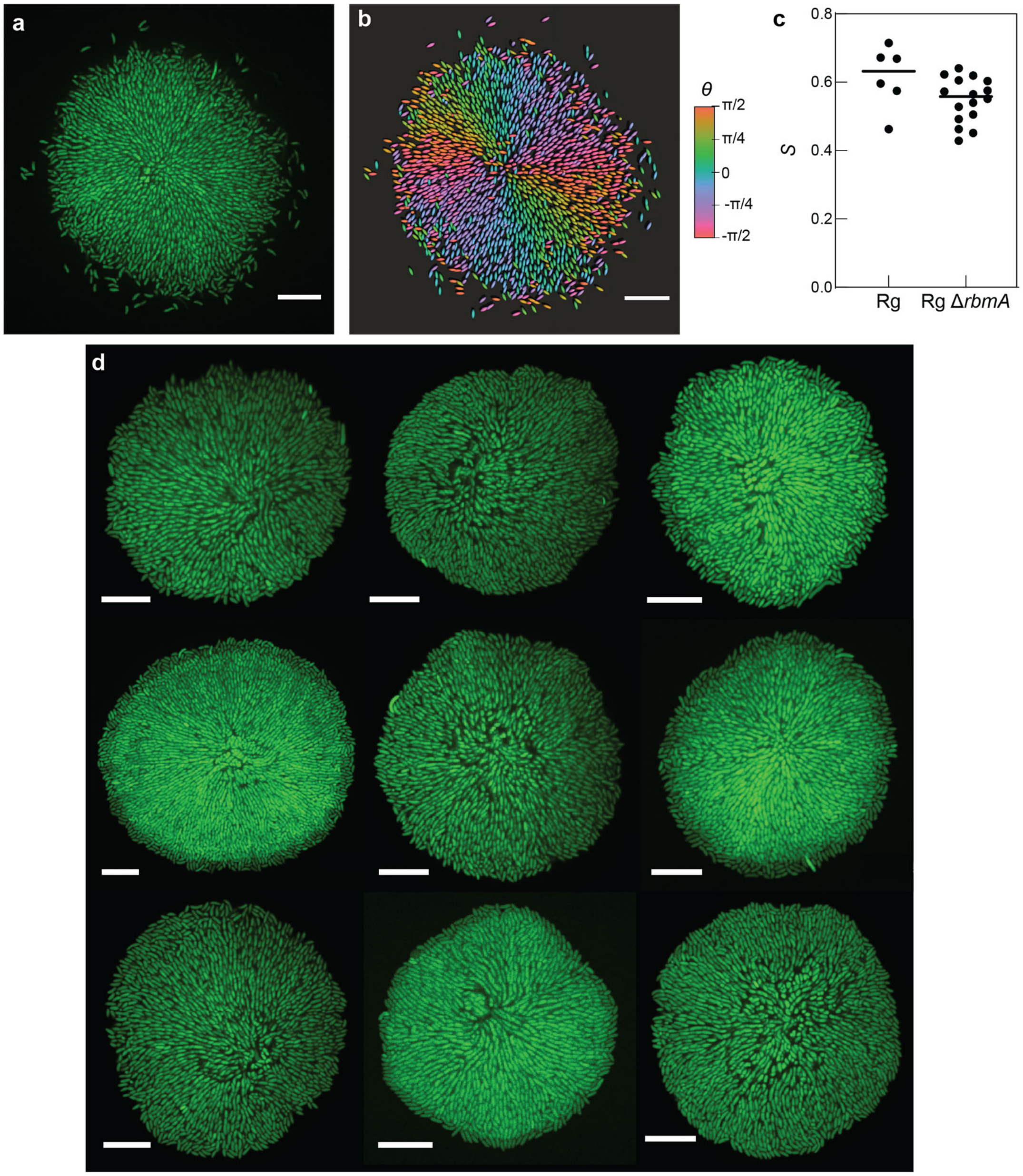

**Extended Data Fig. 1| *V. cholerae* biofilms robustly and reproducibly form aster patterns. a**, Cross-sectional view of the basal plane of a rugose biofilm with an intact *rbmA* gene. **b**, The same biofilm reconstructed where cells are colored based on the angle of the in-plane director. **c**, Radial order parameter $S$ in biofilms formed by the rugose wild-type and Rg$\Delta$*rbmA* strain (labeled as WT* in the main text). We find no statistical difference between the two strains (unpaired *t*-test with Welch's correction) under this growth condition, suggesting that the presence of the cell-to-cell adhesion protein does not interfere with the global organization mechanism. In fact, the rugose wild-type biofilms display the same radial organization even when growing freely on a solid substrate without confinement by the agarose gel, an observation that has been reported[25,26] but never explained. Therefore, we believe that the same macroscopic ordering mechanism applies generally to biofilms grown in any geometry. In this work, we used a confined geometry to increase

the basal area of the biofilm, in order to focus on this emergent organization. **d,** Nine different mature Rg$\Delta$*rbmA* biofilms all exhibiting the aster pattern. The core of the aster, typically defined by the location of the founder cell, does not necessarily coincide with the geometric center of the biofilm due to stochasticity during early biofilm development. Scale bars, 10 μm.

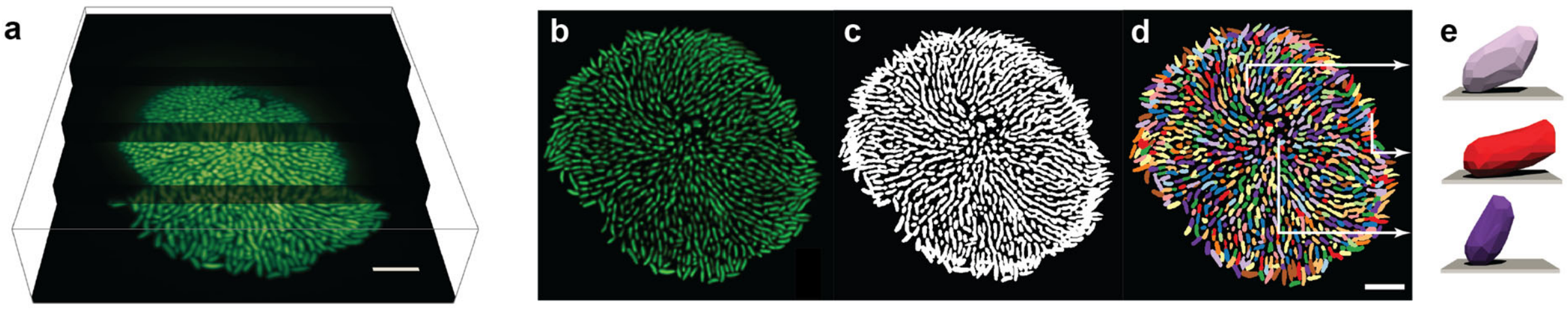

**Extended Data Fig. 2| Cell segmentation procedure. a,** Representative raw 3D image of a WT* biofilm. **b-d**, Intermediate steps during biofilm segmentation; after deconvolution with a measured point spread function (**b**), after binarization (**c**), after segmentation using an adaptive local thresholding algorithm (**d**). **e**, 3D reconstruction of three different segmented cells at three different locations in the biofilm. Scale bars, 10 µm.

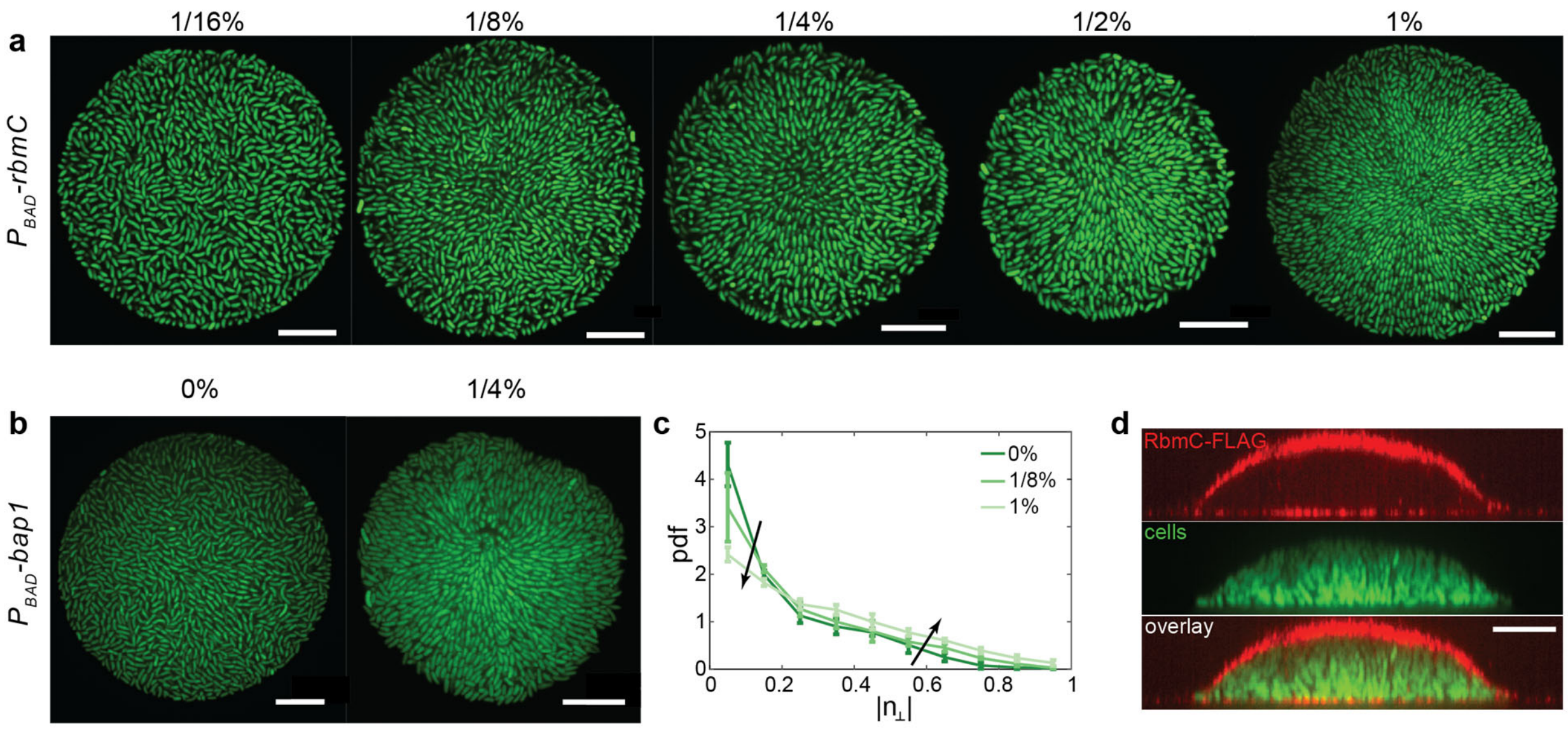

**Extended Data Fig. 3| Cell-to-surface adhesion controls macroscopic cell organization. a**, Cross-sectional view of the basal layer of biofilms with arabinose-inducible *rbmC* expression. In the presence of different concentrations of arabinose, a transition from disorder-to-order was observed with increasing arabinose. **b**, A similar disorder-to-order transition was seen with arabinose-inducible *bap1* expression. Bap1 and RbmC have been shown to contribute to cell-to-surface adhesion in a partially redundant fashion[27]. **c**, Probability distribution function (pdf) of the degree of verticalization $|n_\perp|$ showed a shift to more verticalized cells and less horizontal cells with higher *rbmC* expression controlled by arabinose concentration ($\geq 4$ biofilms; error bars correspond to std). **d**, Spatial distribution of RbmC-3×FLAG stained with Cy3-conjugated anti-FLAG antibodies shows that RbmC localizes on both the glass and gel surfaces. Scale bars, 10 μm.

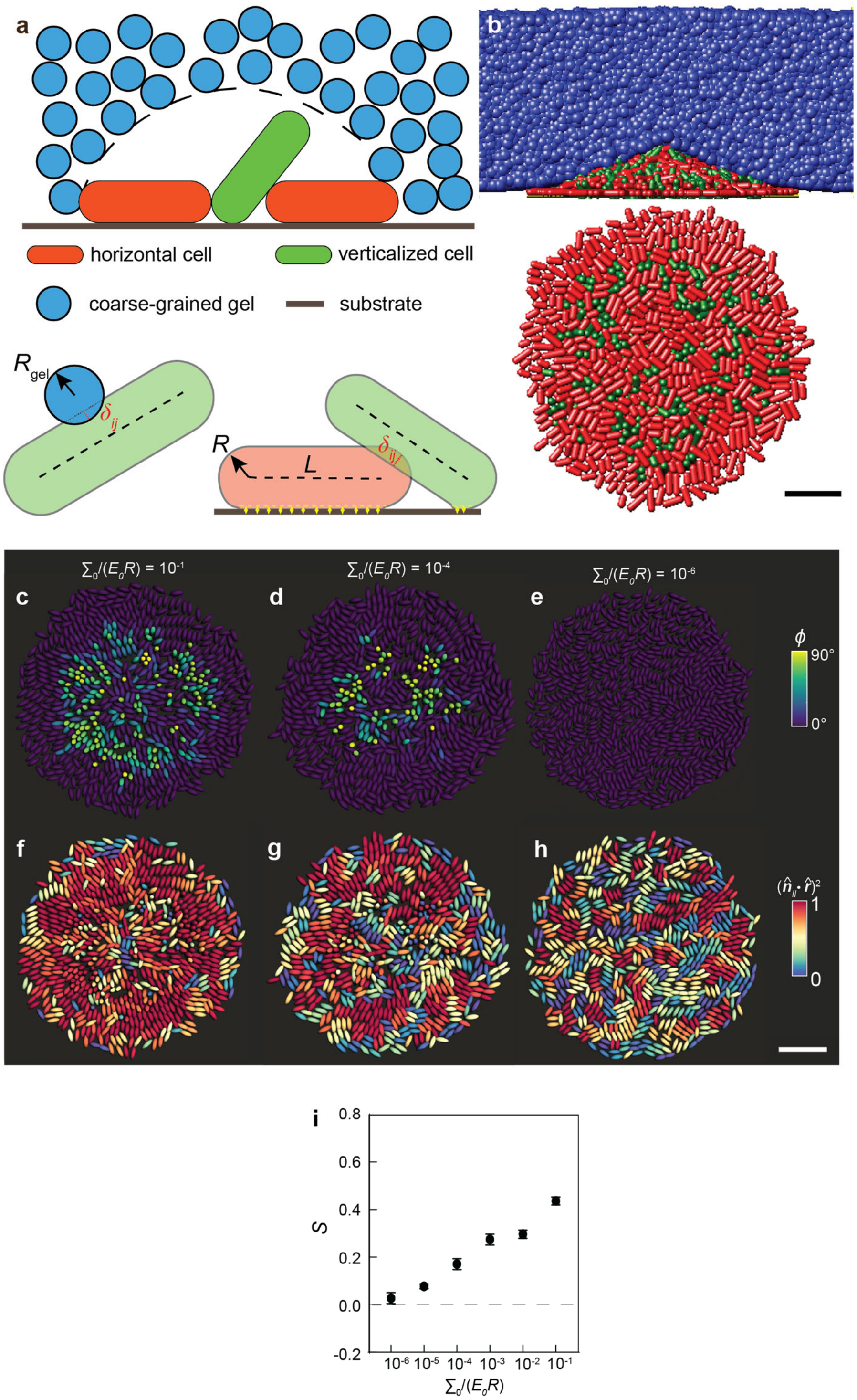

**Extended Data Fig. 4| Agent-based simulations of 3D biofilms with varying adhesion. a**, Schematic illustration of the agent-based simulation consisting of biofilm dwelling cells, which are modeled as growing spherocylinders with length $L(t)$ and radius $R$, and the surrounding hydrogel, which is modeled using a coarse-grained particle system. **b**, Side view (*Top*) and bottom view (*Bottom*) of a representative simulated biofilm. In the bottom view, gel particles are omitted

for clarity. Verticalized cells ($n_\perp > 0.5$) are labelled green, and horizontal cells ($n_\perp \leq 0.5$) are labelled red. Scale bar, 5 μm. **c-e**, Cells in the basal layer, color-coded by the angle $\phi$ each cell makes with the substrate, in biofilms possessing diffferent adhesion strengths. The dimensionless adhesion strength $\tilde{A} = \Sigma_0/RE_0$ is $10^{-1}$, $10^{-4}$, and $10^{-6}$ for **c** to **e**, respectively. **f-h**, The same biofilms as **c-e** with cells color-coded by the degree of radial alignment $(\hat{\boldsymbol{n}}_\parallel \cdot \hat{\boldsymbol{r}})^2$. Scale bar, 10 μm. **i,** Averaged $S$ in biofilms with different dimensionless adhesion strengths. Each data point corresponds to the mean and error bars correspond to the standard deviation of 10 different simulated biofilms.

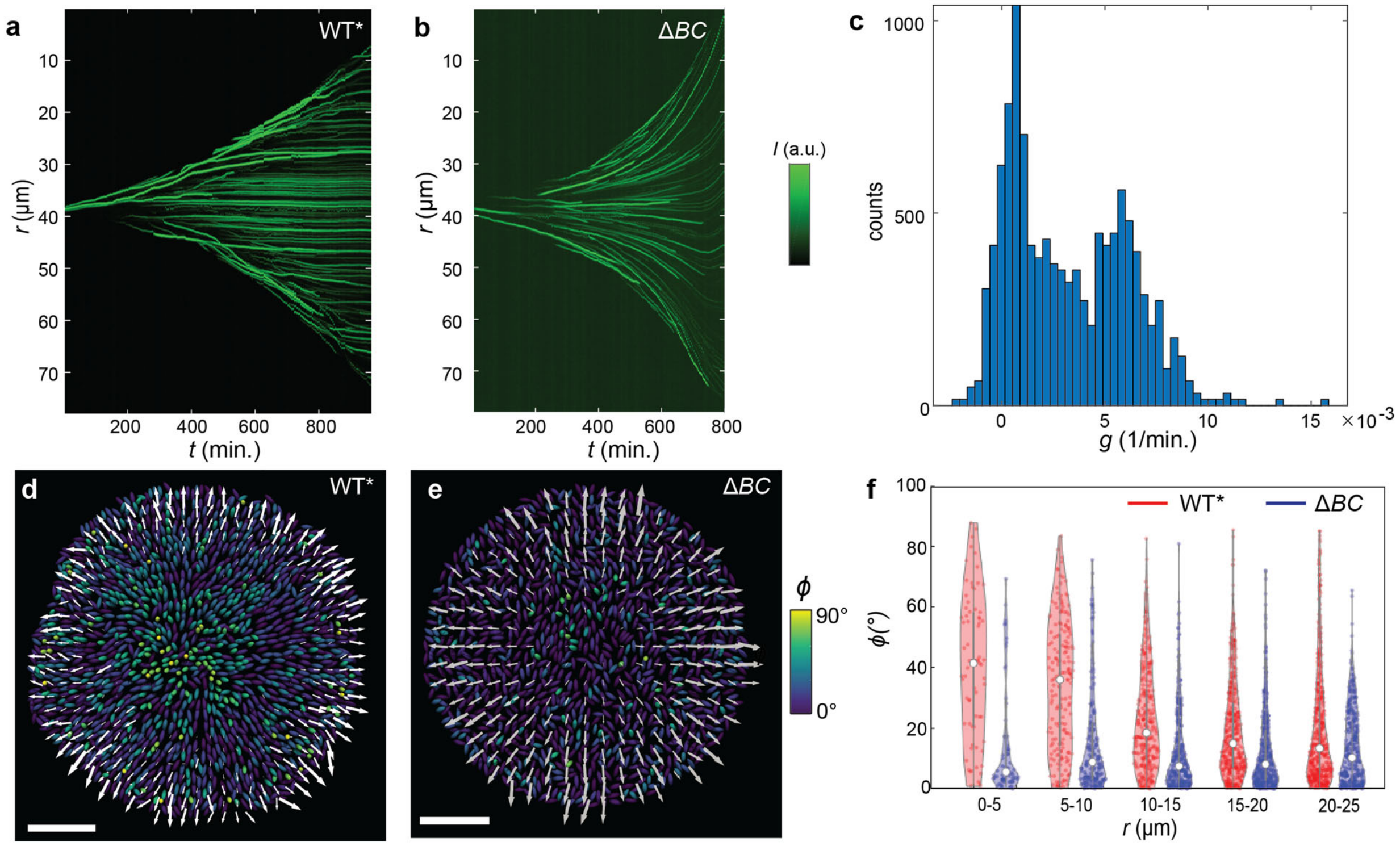

**Extended Data Fig. 5| Biofilms spontaneously develop zones with differential growth due to adhesion-mediated stable verticalization. a, b**, Kymograph of puncta trajectories in a WT* (**a**) and a Δ*BC* mutant biofilm (**b**). In the WT* biofilm a stationary central region develops and expands. In contrast, in the mutant biofilm, no stationary region exists. **c**, Histogram of apparent growth rates showing a bimodal distribution of "growing" and "non-growing" cells. Reconstructed image of a biofilm from a WT* (**d**) and a Δ*BC* mutant biofilm (**e**), where cells are color-coded by the angle they make with respect to the bottom substrate. Overlain are arrows denoting the measured velocity field. **f**, Violin plot showing the distribution of cell orientations at different radii, for the WT* and Δ*BC* mutant biofilms, respectively. The WT* biofilm was able to sustain verticalized cells leading to a growth void in the middle and a nonlinear velocity profile, whereas the Δ*BC* mutant is unable to sustain verticalized cells, therefore leading to a linear velocity profile. Scale bars, 10 μm.

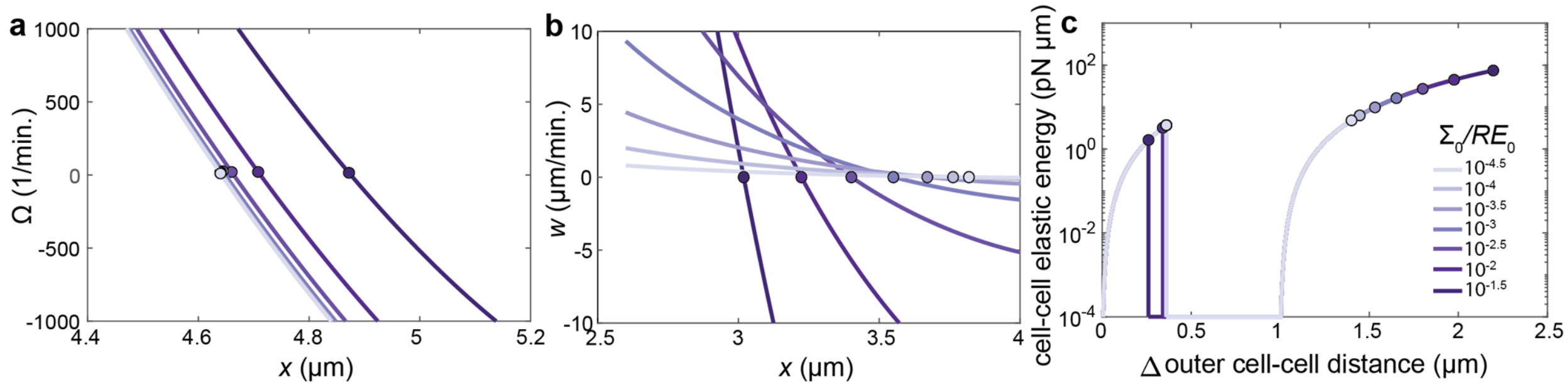

**Extended Data Fig. 6| Comparison of energy landscapes for different adhesion energies. a**, **b**, Determination of the instability point for the verticalization (**a**) and pinch-off instabilities (**b**). Regions in which $\Omega > 0, w > 0$ correspond to regions where fixed points are unstable. Circles denote the transition between stable and unstable behavior. Here $\dot{n}_z = \Omega(x) n_z$ and $\dot{z} = w(x, L/2 + 0.999R)$. **c**, Elastic energy due to cell-to-cell contacts in a cell being squeezed incrementally by two neighbors for different dimensionless adhesion energies, $\widetilde{\mathrm{A}} = \Sigma_0/RE_0$ (see schematic in Fig. 2F). Circles denote the points at which instabilities occur. Here the energy landscape is calculated using the results of the stability analysis.

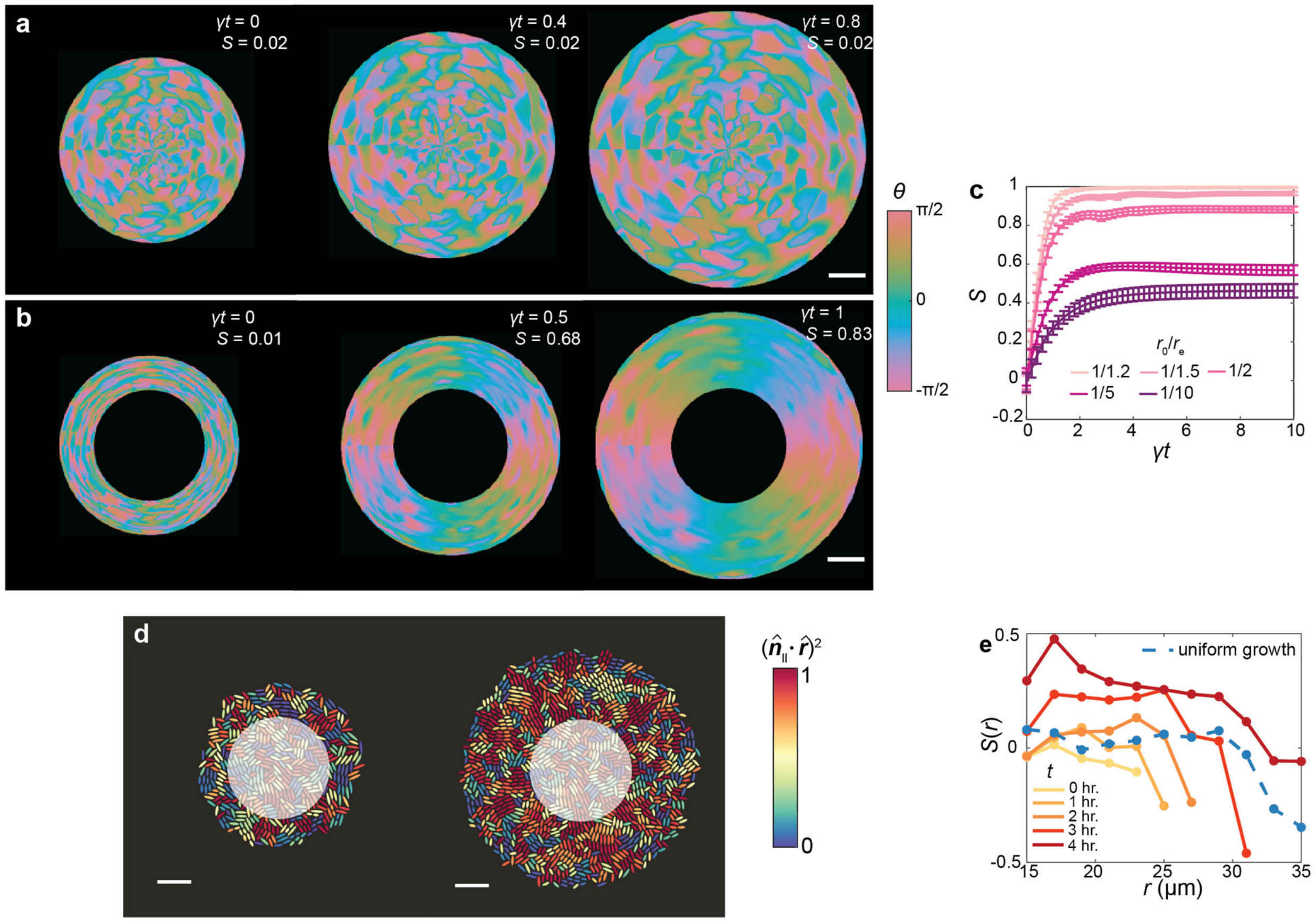

**Extended Data Fig. 7| Continuum modelling and ABSs confirm that a growth void is sufficient to drive radial cell alignment. a**, **b**, Evolution of the director field for a biofilm without (**a**) and with (**b**) a growth void. In panel **b** the radius of the void is $r_0 = (3/5)r_\mathrm{e}$, where $r_\mathrm{e}$ corresponds to the biofilm size when the growth void is introduced. **c**, Evolution of $S$ averaged over the growing region for different dimensionless initial void sizes $r_0/r_\mathrm{e}$. The dimensionless flow alignment parameter $\tilde{L} = \lambda/4q = 1.5$. Each data point corresponds to the mean and error bars correspond to the standard deviation of 10 different random initializations. Radial order emerges over time but plateaus because, unlike the WT* biofilm with an expanding verticalized core, the size of the void is constant in this set of calculations and its effect on cell reorientation is limited to cells near the void due to the $1/r^2$ dependence of the driving force. **d**, Configuration of a 2D biofilm (*left*) right before imposing a growth void and (*right*) 4hr after imposing a growth void (highlighted by the white circle). The color denotes the degree to which the cell orientation coincides with the radial director $(\hat{\boldsymbol{n}}_\parallel \cdot \hat{\boldsymbol{r}})^2$. **e,** Azimuthally averaged $S(r)$ in biofilms after the creation of a growth void at different times. Dashed blue line shows the result from a control biofilm without a growth void. Scale bars, 10 μm.

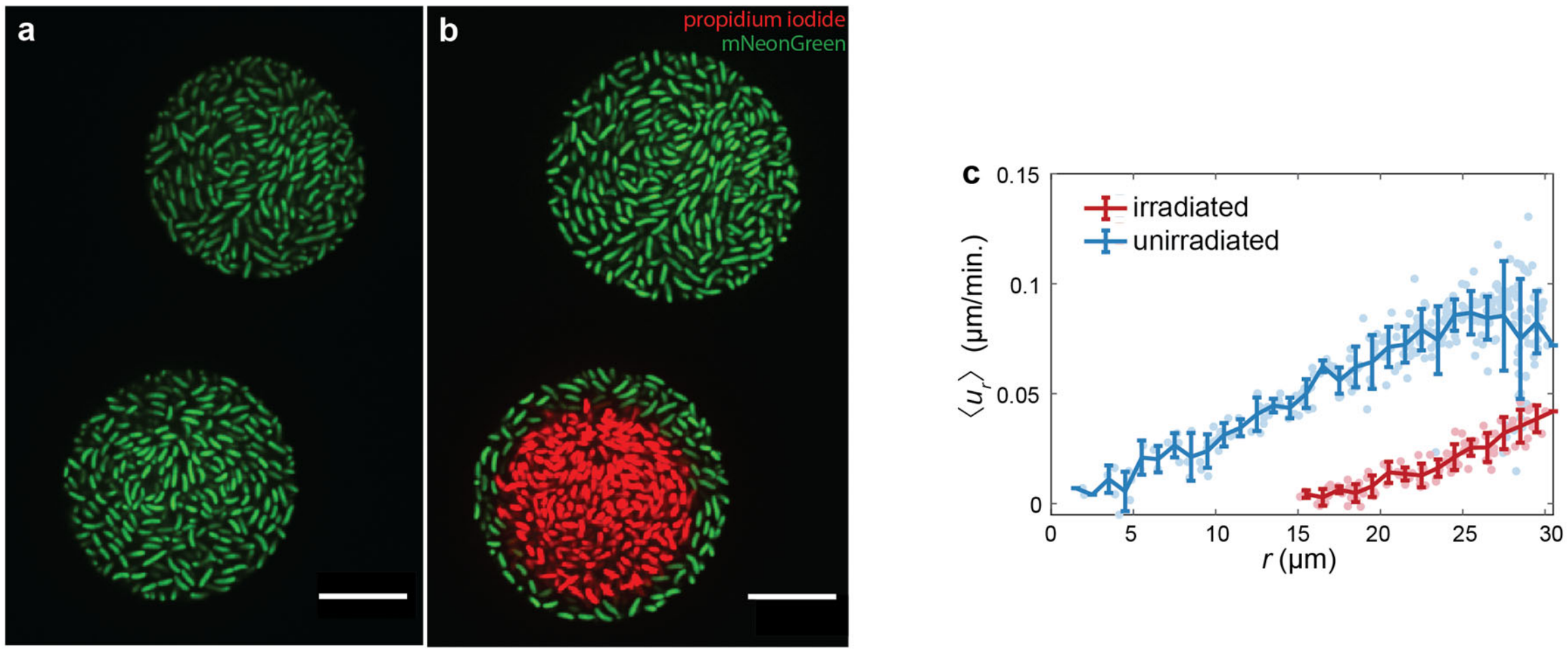

**Extended Data Fig. 8| Laser irradiation causes cell death and results in a growth void. a**, Two $\Delta BC$ biofilms in close proximity before laser irradiation. **b**, The same biofilms after being irradiated in a circular pattern using 405 nm light. Dead cells are stained with propidium iodide. **c**, The measured radial velocity field 1 hour after irradiation (red) and without irradiation (blue). Scale bars, 10 μm.

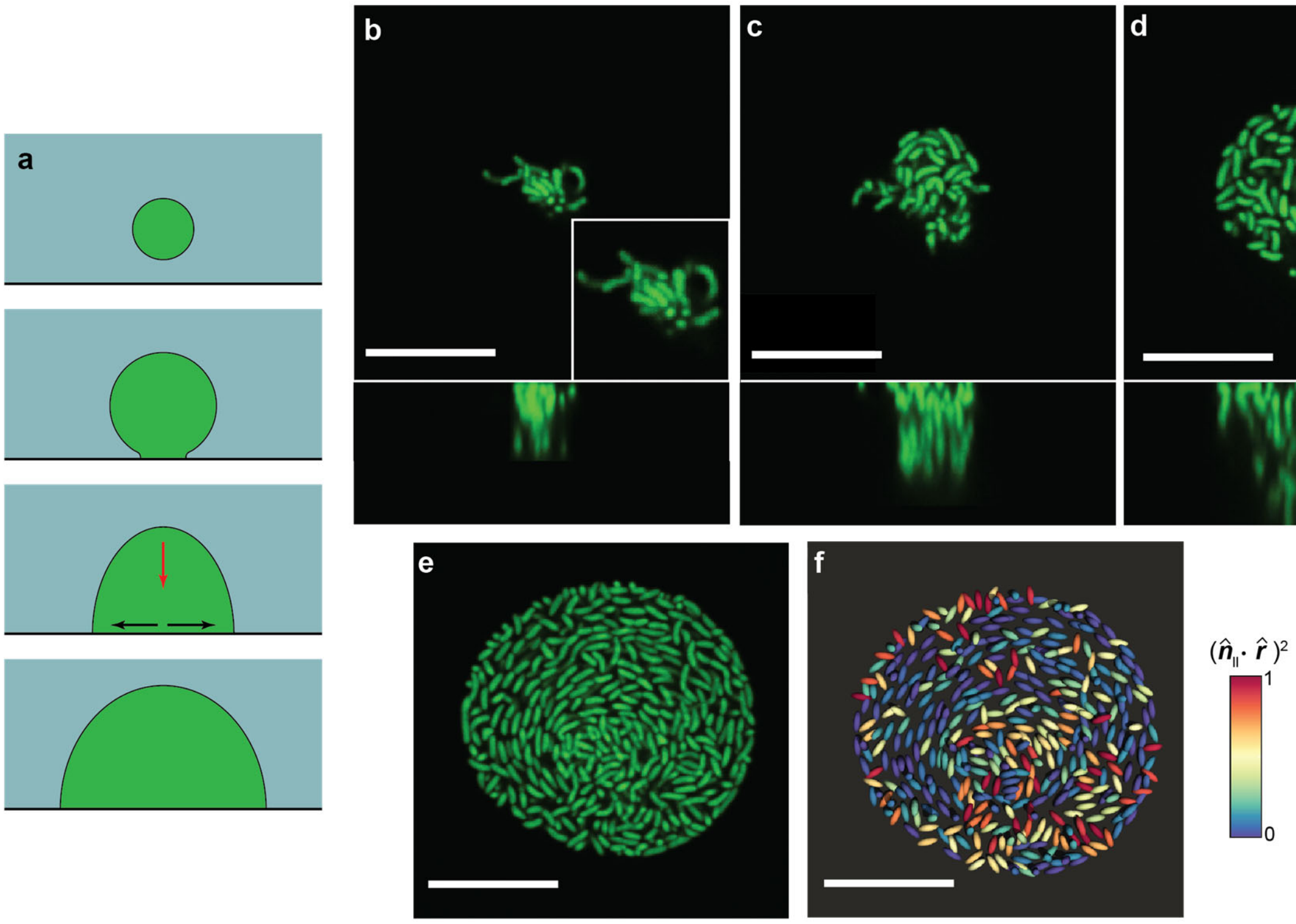

**Extended Data Fig. 9| Nonadherent biofilms have the potential to form vortex structures due to non-uniform growth.** To probe if vortex structures were possible, we imaged $\Delta BC$ biofilms that were initially embedded inside the bulk of the gel instead of starting at the gel-glass interface. This is because, once such a biofilm reached the interface, it preferentially grew at the interface with little increase in biofilm height, leading to excess growth introduced into the basal layer from the bulk, mostly at the center. This process is given schematically in panel **a**. Initially, the biofilm grows completely inside the gel without being in contact with the glass surface. Eventually, the growing biofilm begins to spread between the gel and the glass surfaces and cells grow preferentially along this interface. In this case, excess growth is supplied by cells in the bulk (red arrow) which add to the in-plane growth rate at the center of the basal layer (black arrows). Excess growth at the center gives rise to a negative $f(r,t)$ in Main Text Eq. 2 (see Supplementary Information Section 4 for more discussion), therefore providing a driving force to align cells azimuthally. **b-e**, Time-lapse imaging of a growing $\Delta BC$ biofilm in which a vortex-like structure appears ($t = 6, 9, 12, 15$ hrs). A zoom-in image is provided in **b** when the embedded biofilm first touches the glass surface; at this moment cells are randomly oriented. **f**, Reconstructed biofilm from panel **e** where cells are color-coded based on the degree of radial alignment $(\hat{\boldsymbol{n}}_{\parallel} \cdot \hat{\boldsymbol{r}})^2$. A value of zero denotes cells that are oriented azimuthally (circumferentially).

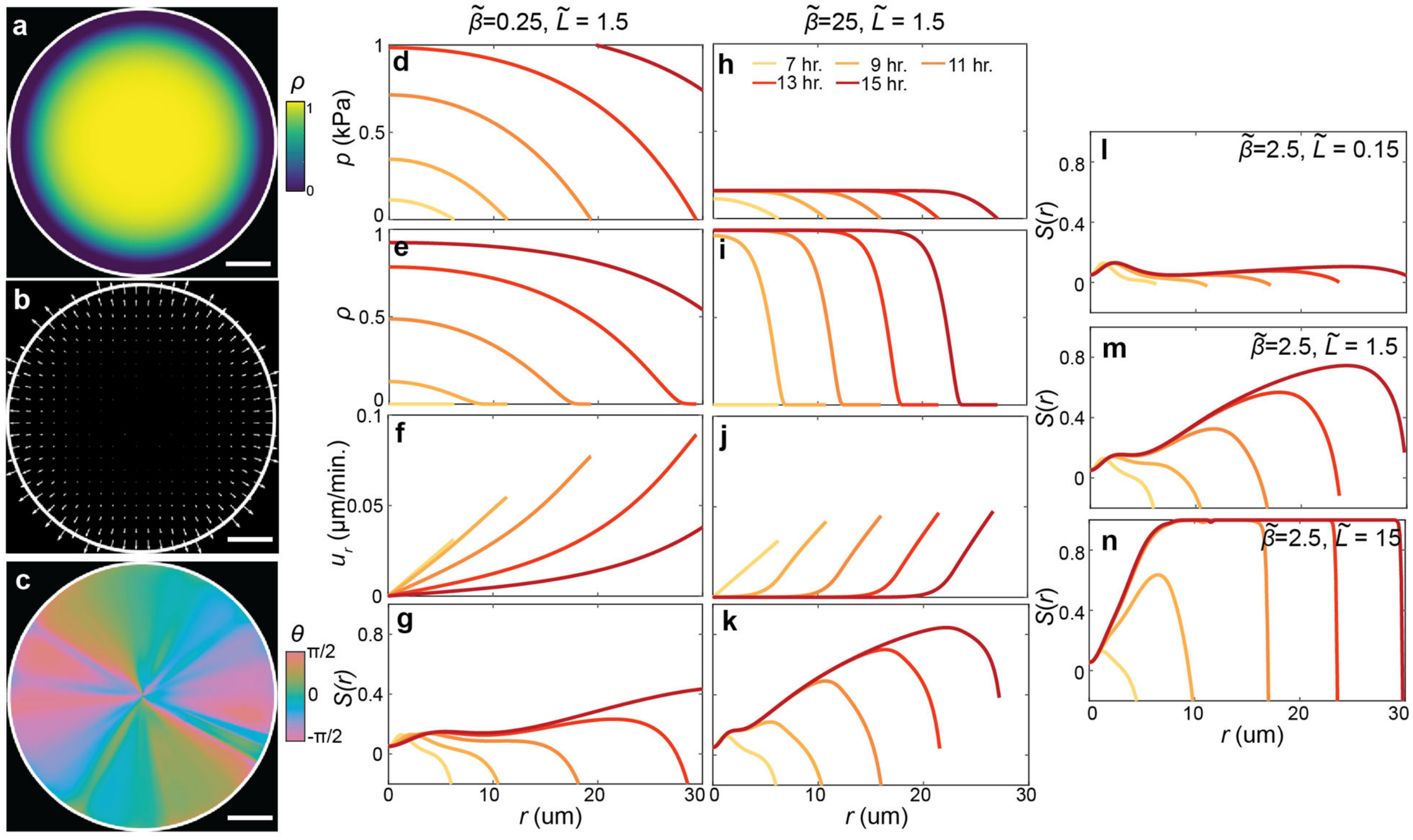

**Extended Data Fig. 10| Representative results from two-phase active nematic model. a-c**, Representative 2D plots of the fraction of verticalized cells $\rho$ (**a**), the velocity field (**b**), and the director field (**c**) for dimensionless verticalization rate $\tilde{\beta} = 2.5$ and dimensionless flow alignment parameter $\tilde{L} = 1.5$. Scale bars, 10 µm. **d-k**, Results of the model for a small verticalization rate $\tilde{\beta} = 0.25$ (**d-g)** and for a large verticalization rate $\tilde{\beta} = 25$ (**h-k)**. **l-n**, Evolution of $S$ for three different flow alignment rates $\tilde{L}$ while keeping the same $\tilde{\beta} = 2.5$. The colors denote model results at different times in panels **d-n**.

## Table S1: List of the strains used in this study

| **Strains** | **Genotype** | **Source** |
|---|---|---|
| JN007 | *vpvc*$^{W240R}$*, ΔrbmA, Δvc1807::Ptac-mNeonGreen*-Spec$^{R}$ | This study |
| JN009 | *vpvc*$^{W240R}$*, ΔrbmA, Δbap1, ΔrbmC, Δvc1807::Ptac-mNeonGreen*-Spec$^{R}$ | This study |
| JN010 | *vpvc*$^{W240R}$*, ΔvpsL, Δvc1807::Ptac-mNeonGreen*-Spec$^{R}$ | This study |
| JN148 | *vpvc*$^{W240R}$*, ΔrbmA, ΔlacZ::Ptac-mNeonGreen-μNS, Δvc1807::Ptac-mScarletI*-Spec$^{R}$ | This study |
| JN150 | *vpvc*$^{W240R}$*, ΔrbmA, Δbap1, ΔrbmC, ΔlacZ::Ptac-mNeonGreen-μNS, Δvc1807::Ptac-mScarletI*-Spec$^{R}$ | This study |
| JN035 | *vpvc*$^{W240R}$*, ΔrbmA, Δbap1, ΔrbmC, Δvc1807::Ptac-mNeonGreen*-Spec$^{R}$, pJY056 | This study |
| JN036 | *vpvc*$^{W240R}$*, ΔrbmA, Δbap1, ΔrbmC, Δvc1807::Ptac-mNeonGreen*-Spec$^{R}$, pJY057 | This study |
| JY489 | *vpvc*$^{W240R}$*, ΔrbmA, rbmC-3×FLAG, Δvc1807::Ptac-mNeonGreen*-Spec$^{R}$ | This study |
| | | |
| **Plasmid** | | |
| pJY056 | Kan$^{R}$, *araC-*$P_{BAD}$*-rbmC* | [26] |
| pJY057 | Kan$^{R}$, *araC-*$P_{BAD}$*-bap1* | [26] |

**Table S2: List of cell parameters used in the agent-based simulations.**

| $R(\mu m)$ | $L_{max}(\mu m)$ | $E_0(\mathrm{Pa})$ | $\gamma(s^{-1})$ | $\Sigma_0(\mathrm{Nm}^{-1})$ | $\eta_0(\mathrm{Pa \cdot s})$ | $\eta_1(\mathrm{Pa \cdot s})$ |
|---|---|---|---|---|---|---|
| 0.8 | 3.6 | 300 | $3.12 \times 10^{-4}$ | $7.3 \times 10^{-6}$ | 20 | $2 \times 10^5$ |

**Table S3: List of gel parameters used in the agent-based simulations.**

| $R_{\mathrm{gel}}(\mu\mathrm{m})$ | $E_1(\mathrm{Pa})$ | $k_{\mathrm{r}}\ (\mathrm{Nm}^{-1})$ | $k_{\zeta}(\mathrm{Nm}^2)$ | $\xi_0(\mu\mathrm{m})$ | $\zeta_0(°)$ | $\Sigma_1(\mathrm{Nm}^{-1})$ |
|---|---|---|---|---|---|---|
| 1.0 | 1500 | $6 \times 10^{-3}$ | $6.6 \times 10^{-4}$ | 0.6 | 120 | $6 \times 10^{-2}$ |

**Movie captions**

**Movie S1:** A time-lapsed cross-sectional view of the basal plane of a growing WT* biofilm. Imaging began immediately after seeding the founder cell. The total duration of the movie is 20 hr. The scale bar is 10 μm.

**Movie S2:** A time-lapsed cross-sectional view of the basal plane of a WT* biofilm with mNeonGreen labelled puncta. Imaging began 8 hr. after seeding the founder cell. The total duration of the movie is 16 hr. The scale bar is 10 μm.

**Movie S3:** Representative results of the agent-based simulations. (*Left*) Bottom view of a biofilm in which cells are colored red if horizontal ($n_{\perp} \leq 0.5$) or green if vertical ($n_{\perp} > 0.5$). (*Right*) Side view of the same biofilm (red/green) and surrounding course-grained gel particles (blue). The duration of the movie is 18 hr.

**Movie S4:** Representative results of a quasi-2D agent-based simulation in which a growth void is introduced at the center of a growing biofilm. The movie is slowed down 1.5-fold after the introduction of the growth void. The total duration of the movie is 16 hr.